%% file: main.tex
\documentclass[preprint,12pt]{elsarticle}

\usepackage[T1]{fontenc}
\usepackage[utf8]{inputenc}
\usepackage{textcomp}
\usepackage{graphicx}
\usepackage{amsmath}
\usepackage{booktabs}
\usepackage{array}
\usepackage{url}
\usepackage{longtable}
\usepackage{etoolbox}
\AtBeginEnvironment{table}{\linespread{1}\selectfont}

\graphicspath{{figs/}}
\biboptions{numbers,sort&compress}

\journal{Applied Thermal Engineering}

\begin{document}

\begin{frontmatter}

\title{Binder migration across film consolidation and pore emptying in
electrode drying: a continuous model validated against drying-rate,
thickness, and schedule experiments}

\author[kier,ust]{Younghwan Joo\corref{cor}}
\ead{yhjoo@kier.re.kr}
\author[kier]{Sanghyun Oh}
\author[kier,ust]{Sung-il Kim\corref{cor}}
\ead{praygod@kier.re.kr}
\cortext[cor]{Corresponding authors.}
\affiliation[kier]{organization={Energy Efficiency Research Division,
  Korea Institute of Energy Research},
  addressline={152 Gajeong-ro, Yuseong-gu},
  city={Daejeon},
  postcode={34129},
  country={Republic of Korea}}
\affiliation[ust]{organization={Energy Engineering,
  University of Science \& Technology},
  addressline={217 Gajeong-ro, Yuseong-gu},
  city={Daejeon},
  postcode={34113},
  country={Republic of Korea}}

\begin{abstract}
Drying battery electrodes faster saves energy but drives dissolved binder
toward the film surface, depleting the current-collector interface and
weakening adhesion. Models of this migration stop convecting the binder when
film consolidation ends, and models of the subsequent pore emptying omit the
binder or start from an assumed binder field, so no simulation has followed
binder convection through both stages that a drying schedule acts on. This
study
presents a continuous simulation of the dissolved binder field across both
stages, driven by the published drying rates and completed by effective
descriptions of the two questions the base transport physics leaves open:
where the liquid of the consolidating film carries the binder, and when the
binder stops moving as the pores empty. With a single parameter set, the
simulation reproduces three types of published measurements: the change of
the surface-to-interface binder ratio with drying rate over a factor of 5.6
and the surface and interface binder levels of films six times thicker;
concentration profiles through the film interior; and the response of the
interface binder to the time at which the drying rate is switched
mid-process. Applied without further adjustment to a published three-stage
schedule, the model reproduces the shortened drying and the full interface
recovery when the rate is lowered early, and shows that its final high-rate
stage begins after the binder has stopped. Combined with a coater-scale
process model, the
model can be used to identify operating points for electrode production
that improve energy efficiency and product quality together.
\end{abstract}

\begin{keyword}
Electrode drying \sep Binder migration \sep Lithium-ion battery
manufacturing \sep Porous media \sep Drying schedule \sep Continuum
modeling
\end{keyword}

\end{frontmatter}

\input{sections/nomenclature}

\input{sections/01_introduction}
\input{sections/02_model}
\input{sections/03_data_calibration}
\input{sections/04_results}
\input{sections/05_discussion}
\input{sections/06_conclusions}

\section*{CRediT authorship contribution statement}
\textbf{Younghwan Joo}: Conceptualization, Methodology, Software,
Validation, Formal analysis, Investigation, Data curation, Writing --
original draft, Writing -- review \& editing, Visualization.
\textbf{Sanghyun Oh}: Resources, Writing -- review \& editing.
\textbf{Sung-il Kim}: Supervision, Project administration, Funding
acquisition, Writing -- review \& editing.

\section*{Declaration of competing interest}
The authors declare that they have no known competing financial interests
or personal relationships that could have appeared to influence the work
reported in this paper.

\section*{Acknowledgements}
This work was conducted under the framework of the Research and
Development Program of the Korea Institute of Energy Research (KIER)
(C6-2410-34).

\section*{Data availability}
The numerical data underlying the figures of this study and the measured
values read from the published figures used for comparison are available
at \url{https://doi.org/10.6084/m9.figshare.34704411}.

\section*{Declaration of generative AI and AI-assisted technologies in the manuscript preparation process}
During the preparation of this work the authors used Claude (Anthropic)
to assist with writing the manuscript and with the analysis and plotting
work. After using this tool, the authors reviewed and edited the content as needed and take
full responsibility for the content of the published article.

\bibliographystyle{elsarticle-num}
\bibliography{refs}

\end{document}

%% file: sections/nomenclature.tex
\section*{Nomenclature}

\noindent
\begin{tabular}{@{}p{2.2cm}p{11.0cm}@{}}
\multicolumn{2}{@{}l}{\textit{Roman symbols}}\\
$a$ & absolute-scale factor of the profile comparison, \S\ref{sec:results-endpoint} (--)\\
$B$ & dimensionless group of the pumping-failure closure, Eq.~\eqref{eq:model-sfail} (--)\\
$c_b$ & dissolved binder concentration, per unit total film volume (kg\,m$^{-3}$); $c_{b0}$, initial value\\
$d$ & mixing depth of the calendering operator, Eq.~\eqref{eq:model-calender} (m)\\
$\hat c$ & normalized binder concentration, $c_b/c_{b0}$ (--)\\
$\hat c_i$, $\hat c_s$ & interface and surface values of $\hat c$ (--)\\
$D_0$ & effective binder diffusivity prefactor (m$^2$\,s$^{-1}$)\\
$D_b$ & effective binder diffusivity (m$^2$\,s$^{-1}$)\\
$\mathcal D$ & binder diffusivity scale, $D_0\,\phi_l^{1/2}$ (m$^2$\,s$^{-1}$)\\
$f$ & localization fraction of the pumping profile (--)\\
$f_1(S)$ & binder immobilization function, Eq.~\eqref{eq:model-f1} (--)\\
$F(\mathrm{Pe})$ & surface enrichment $\ln(c_b^{\,\mathrm{surf}}/c_b^{\,\mathrm{int}})$ as a function of the P\'eclet number, Eq.~\eqref{eq:model-pe} (--)\\
$F_b$ & effective binder-advecting flux, Eq.~\eqref{eq:model-binder} (m\,s$^{-1}$)\\
$F_l$ & superficial liquid flux (m\,s$^{-1}$)\\
$g$ & binder flux at the receding surface, $\gamma\,c_b$ (kg\,m$^{-2}$\,s$^{-1}$)\\
$G$ & normalized binder flux in the scaled coordinate, Eq.~\eqref{eq:model-collapse} (m\,s$^{-1}$)\\
$H$ & film thickness (m); $H_0$, initial value\\
$J$ & Leverett function evaluated at intermediate saturation (--)\\
\end{tabular}

\medskip
\noindent
\begin{tabular}{@{}p{2.2cm}p{11.0cm}@{}}
$k$ & rate sensitivity of the surface enrichment, slope of $\ln(c_b^{\,\mathrm{surf}}/c_b^{\,\mathrm{int}})$ against $\dot m$ (m$^2$\,s\,g$^{-1}$); $k_{70}$, $k_{400}$, values of the 70 and 400~\textmu m series\\
$\dot m$ & prescribed evaporative mass flux (g\,m$^{-2}$\,s$^{-1}$)\\
$m$ & concentration--viscosity exponent (--)\\
$n$ & thickness exponent, $\ln(k_{400}/k_{70})/\ln(H_{400}/H_{70})$ (--)\\
$p$ & percolation exponent (--)\\
$p_c$, $p_g$, $p_l$ & capillary, gas and liquid pressures (Pa)\\
$p_{c0}$ & capillary pressure scale, $\sigma\sqrt{\varepsilon/\kappa}$ (Pa)\\
$\mathrm{Pe}$ & P\'eclet number, $\gamma H/\mathcal D$ (--)\\
$q$ & immobilization exponent (--)\\
$S$ & liquid saturation (--)\\
$S_c$ & arrest point of the immobilization function (--)\\
$S_{\mathrm{fail}}$ & pumping-failure threshold saturation (--)\\
$S_{\mathrm{il}}$ & irreducible liquid saturation (--)\\
$S_M$ & macropore-network saturation (--)\\
$S_{\mathrm{perc}}$ & percolation constant of the closure (--)\\
\end{tabular}

\noindent
\begin{tabular}{@{}p{2.2cm}p{11.0cm}@{}}
$t$ & time (s); $t_{50}$, response midpoint; $t_{\mathrm{trans}}$, switching time; $t_{c,1}$, $t_{c,2}$, characteristic transition times reported in the switch-time experiments\\
$u_b$, $u_l$ & binder and liquid velocities (m\,s$^{-1}$)\\
$u(\xi)$ & convection velocity relative to the scaled grid, Eq.~\eqref{eq:model-collapse} (m\,s$^{-1}$)\\
$w_n$ & nanopore reservoir volume fraction (--)\\
$x$ & through-film coordinate (m)\\
\end{tabular}

\medskip
\noindent
\begin{tabular}{@{}p{2.2cm}p{11.0cm}@{}}
\multicolumn{2}{@{}l}{\textit{Greek symbols}}\\
$\alpha$ & partial homogenization coefficient of the calendering observation operator (--)\\
$\beta_b$ & carrier coupling factor (--)\\
$\gamma$ & surface recession rate, $\dot m/\rho_l$ (m\,s$^{-1}$)\\
$\delta$ & accumulation-layer thickness, $\mathcal D/\gamma$ (m)\\
$\Delta H_{\mathrm{cal}}$ & measured calendering compression (m)\\
$\varepsilon$ & film porosity (--)\\
$\kappa$ & intrinsic permeability (m$^2$)\\
$\kappa_{\mathrm{rl}}$ & relative permeability of the liquid (--)\\
$\kappa_{\mathrm{rl},M}$ & relative permeability of the macropore carrier network (--)\\
$\lambda$ & depth enrichment rate, relative increase of the binder concentration per unit depth, $u/\mathcal D$ (m$^{-1}$)\\
$\mu_l$ & liquid viscosity (Pa$\cdot$s); $\mu(\hat c_s)$, concentration-dependent value\\
\end{tabular}

\medskip
\noindent
\begin{tabular}{@{}p{2.2cm}p{11.0cm}@{}}
$\xi$ & scaled through-film coordinate, $x/H$ (--)\\
$\xi^{*}$, $\xi_p$ & dual-mode shrinkage parameters (--)\\
$\rho_l$ & liquid density (kg\,m$^{-3}$)\\
$\sigma$ & surface tension (N\,m$^{-1}$)\\
$\phi_l$, $\phi_s$ & liquid and solid volume fractions of the film (--); $\phi_{s0}$, initial solid fraction\\
$\phi$ & liquid volume fraction as defined in the source experiments: volume of liquid over volume of liquid and solids, air not counted (--)\\
$\phi_{t1}$, $\phi_{t2}$ & schedule switching thresholds, in $\phi$ (--)\\
$\phi_{c1}$ & first characteristic value of $\phi$ reported in the switch-time experiments of the three-stage study (--)\\
$\Phi(\xi)$ & localized pumping profile, Eq.~\eqref{eq:model-phi} (--)\\
$\chi$ & prefactor of the pumping-failure closure (--)\\
$\Psi(\xi)$ & dual-mode shrinkage kinematics, Eq.~\eqref{eq:model-psi} (--)\\
\end{tabular}

%% file: sections/01_introduction.tex
\section{Introduction}
\label{sec:intro}

Binder migration during electrode drying is by now a well-characterized
failure mode. As solvent evaporates from a slot-die-coated anode, the liquid
flow toward the surface carries dissolved PVDF with it; the film surface is
enriched and the interface with the current collector is depleted
\cite{muller.etal2017InvestigationBinder}, and peel
strength rises monotonically with the binder concentration at the fracture
plane \cite{jaiser.etal2016InvestigationFilm}. Since migration
strengthens with drying rate, the phenomenon directly affects the main
trade-off of dryer operation: drying faster saves energy and line length
\cite{susarla.etal2018ModelingAnalysis} but weakens adhesion. Infrared and
hot-air combined drying, for example, raises the drying rate and has been
estimated to save 28--42\% of the drying energy of NMP-based electrodes
compared with hot-air drying alone, while its authors name binder migration
as the quality limit of faster drying \cite{guo.etal2026EnergySaving}. What is
less settled is what a simulation must deliver before it can be trusted for setting drying conditions around this trade-off. We take the following as the
requirement. A drying oven has three operating variables: drying intensity
(temperature, air flow), coating amount (wet thickness and loading), and
the placement of drying intensity along the oven (zone scheduling). A
simulation supports the design of drying conditions only if a single parameter set responds to all
three variables with the correct direction and magnitude. A model that is
re-tuned for each variable is a correlation, not a tool.

That requirement can be tested today, without new experiments, because for
the NMP--PVDF graphite system the literature already contains one suitable
data set per variable. For the drying-rate variable,
\cite{muller.etal2017InvestigationBinder} provides four drying rates
spanning a factor of 5.6 on 70~\textmu m anodes with the
surface-to-interface binder ratio measured at each rate. For the loading
variable, the same study extends to 400~\textmu m coatings, six times the
thickness, and resolves binder cross-sections through the film. For the scheduling variable, \cite{jaiser.etal2016InvestigationFilm}
measured the response of interface binder and adhesion to the time at
which the drying rate is switched from high to low, resolving a sigmoidal
transition, and \cite{jaiser.etal2017DevelopmentThreestage} demonstrated
a three-stage schedule that restores adhesion while shortening drying by
40\%. Together the three data sets provide three independent validation
cases for any candidate model: each isolates a different operating
variable, and, as shown later, agreement with one data set does not imply
agreement with another.

Existing models describe the convection of the binder in only part of the
electrode drying process, and the reason is structural. Electrode drying
proceeds in two stages. During film consolidation the particle network
shrinks and expels solvent while the film remains saturated; during pore
emptying the network is fixed and the liquid retreats into its pore space.
The model families divide along the same boundary. Binder-transport models
convect the binder only in the consolidation stage: the model of Font et
al.~\cite{font.etal2018BinderMigration} ends the calculation when the pores
begin to empty, and in the model of Zihrul et
al.~\cite{zihrul.etal2023ModelDevelopment} convection ends when the film
stops shrinking, after which the binder only diffuses and becomes immobile
as the pores dry out.
Models of the pore-emptying stage were developed for heat and solvent
transport and do not include dissolved binder
\cite{susarla.etal2018ModelingAnalysis,kumberg.etal2021InfluenceLayer}.
The pore-network model of Wang et
al.~\cite{wang.etal2026NumericalSimulation} also covers pore emptying alone:
it transports conductive-additive particles from a uniform initial field,
and its authors note that a gradient of binder and particles formed during
consolidation may change the result. A
concurrent pore-scale study \cite{weichel.etal2026PorescaleModeling}
resolves two-phase flow on a real electrode geometry, again for pore
emptying only, from a uniform initial binder field, and without
quantitative comparison to experiment. Because no family convects the
binder across the stage boundary, no existing model can follow a drying
schedule across it. Time-varying drying rates have been simulated within
film consolidation alone, by the continuum model of Font et
al.~\cite{font.etal2018BinderMigration} and by the three-dimensional
particle-based model of Lombardo et
al.~\cite{lombardo.etal2021CarbonbinderMigration} for a cathode slurry, which
reproduced qualitatively the benefit of a low-rate stage in the middle of
drying shown by the three-stage schedule of
\cite{jaiser.etal2017DevelopmentThreestage}. Particle-based models of this
kind end when the particles stop moving and therefore correspond mainly to
film consolidation \cite{wang.etal2026NumericalSimulation}. In the
switch-time experiments, however, the times at which the drying rate is
switched fall on both sides of the stage boundary, so a model confined to
one stage cannot reproduce the full response. To the best of our knowledge, no model
has been compared with all three data sets at the same time.

In this study, the model is built to be compared with all three data sets
at once. A single continuous equation for the dissolved binder
concentration is solved across film consolidation and pore emptying,
driven by a prescribed evaporative flux, that is, by the measured drying
rate imposed directly as the rate at which solvent leaves the film surface.
Prescribing the evaporative flux is the only available way to reproduce the
literature experiments, whose air-side conditions are unpublished, and the
evaporative flux is also the variable through which the electrode model
connects to the coater-scale process model of Joo et
al.~\cite{joo.etal2026StatedependentTransient}. Three contributions
follow. First, a continuous binder model for the entire drying process,
closed by a small set of corrective physics terms, each introduced to
reproduce a specific observation. Second, a single parameter set that
reproduces all three data sets, reported with an explicit separation of
what was fitted from what was predicted. Third, the finding that the
switch-time experiment determines a quantity that measurements on dried
films cannot: the saturation at which binder motion ends during pore
emptying. A dried film records only where the binder finally came to rest, whereas
switching the drying rate at different times shows when the binder stops
responding to it; the value calibrated on the switch-time experiment is
consistent with the saturation range below which the source experiments
indicate that the binder no longer moves. Finally, the scope of this study is limited to the
electrode side rather than the whole coater oven; oven-level operating
variables enter only through the prescribed evaporative flux, and their
coupling, toward a tool for designing operating conditions that evaluates oven energy and interface quality in one calculation, is discussed as future work.

%% file: sections/02_model.tex
\section{Model}
\label{sec:model}

\subsection{Governing equations and the stage transition}
\label{sec:model-governing}

\begin{figure}
  \centering
  \includegraphics[width=\linewidth]{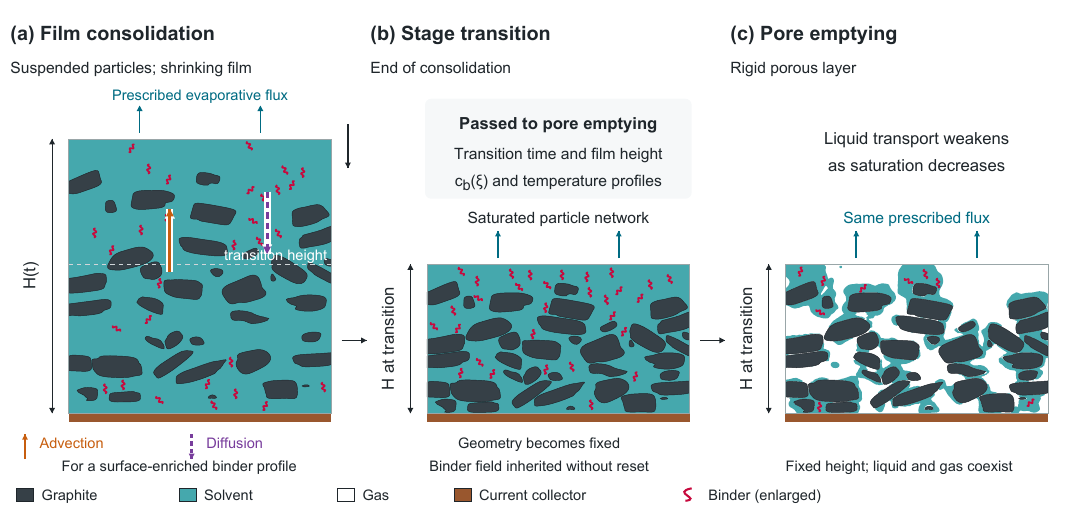}
  \caption{Film consolidation, stage transition and pore emptying in
section: (a) shrinking solvent-filled film with suspended particles;
(b) saturated particle network at the end of consolidation; (c) liquid and
gas in the rigid network at fixed height.}
  \label{fig:model-overview}
\end{figure}

The physical picture the model discretizes is the following. While the film
consolidates, the particle network is still contracting: evaporation removes
solvent at the surface, and wherever particles move closer together the
interstitial liquid is squeezed out and pumped toward the surface, carrying
dissolved binder with it. Binder diffusion cannot keep pace with the surface
loss, so the displaced binder accumulates near the top of the film instead
of redistributing. Once the network becomes rigid, drying continues by emptying the
pore space: menisci retreat into the film, liquid transport weakens as
saturation falls, and at some point the binder stops moving altogether. The
binder concentration left at the current-collector interface is decided by
the competition between these processes, which is why the binder field must
cross the stage boundary without interruption. Fig.~\ref{fig:model-overview} sketches
the two stages and what passes between them.

In both stages the dissolved binder concentration $c_b(x,t)$, the mass of
binder in solution per unit total film volume (kg\,m$^{-3}$), obeys the
same one-dimensional advection--diffusion equation, expressed in
conservative form as follows:
\begin{equation}
\frac{\partial c_b}{\partial t}
+ \frac{\partial}{\partial x}\!\left[
  -\frac{D_{b}}{\phi_l}\frac{\partial c_b}{\partial x}
  + \frac{F_b}{\phi_l}\,c_b \right] = 0,
\label{eq:model-binder}
\end{equation}
where $x$ is the through-film coordinate measured from the current
collector, $t$ is time, $\phi_l$ is the liquid volume fraction, $D_b$ is
the effective binder diffusivity, and $F_b$ is the effective
binder-advecting flux, the superficial flux (volume of liquid per unit
film area and time) at which binder is carried through the film. $F_b$
equals the superficial liquid flux $F_l$ in the base model and is modified
away from it only by the corrective terms of \S\ref{sec:model-corrective},
which act on the binder alone and leave the liquid balance untouched.
Eq.~\eqref{eq:model-binder} follows from the binder mass balance in the
liquid phase: writing $c_l = c_b/\phi_l$ for the concentration per unit
liquid volume, the balance
$\partial(\phi_l c_l)/\partial t + \partial_x(F_b\,c_l - D_b\,\partial_x c_l) = 0$
reduces to Eq.~\eqref{eq:model-binder} wherever $\phi_l$ is spatially
uniform, which holds exactly during consolidation; in pore emptying the
same form is kept with $\phi_l = \varepsilon S$, where $\varepsilon$ is the
film porosity and $S$ the liquid saturation, neglecting the
saturation-gradient contribution to the diffusive flux, a simplification
absorbed in the effective diffusivity. The conserved quantity is accordingly the
binder inventory per unit area, $\int c_b\,dx$, and the concentration
definition does not change at the stage transition. The two stages differ
in the domain and in how $\phi_l$, $D_b$ and $F_b$ are built.

During film consolidation the film is saturated and the domain shrinks with
the film, $0 \le x \le H(t)$. Drying is imposed as a prescribed evaporative
mass flux, so the surface recedes at
\begin{equation}
\frac{dH}{dt} = -\gamma, \qquad \gamma = \frac{\dot m}{\rho_l},
\label{eq:model-height}
\end{equation}
where $H(t)$ is the film thickness, $\gamma$ is the rate at which the
surface recedes, $\dot m$ is the prescribed evaporative mass flux (mass of
solvent evaporated per unit film area and time), and $\rho_l$ is the
liquid density. Solvent conservation with a spatially uniform solid
fraction, $\phi_s = \phi_{s0}H_0/H$, where $\phi_{s0}$ and $H_0$ are the
initial solid fraction and film thickness, \sloppy yields the superficial
liquid flux $F_l = \gamma\,\phi_s\,x/H$, which the base model uses directly
as the binder-advecting flux, $F_b = F_l$. The diffusivity follows a
Bruggeman closure, $D_b = D_0\,\phi_l^{3/2}$, where $D_0$ is the effective
binder diffusivity prefactor, the one transport coefficient of the model
that is fitted. At the receding surface the binder flux satisfies the
exact no-outflow condition relative to the moving boundary,
$j_b - \dot H\,c_b = 0$ at $x = H(t)$, where $j_b$ is the total (advective
plus diffusive) binder flux, positive toward the surface, and $\dot H =
dH/dt$; with $\dot H = -\gamma$ this is the surface condition $g =
\gamma\,c_b$ of the implementation, where $g$ is the binder flux at the
receding surface. Binder cannot leave the film, only redistribute, and the
inventory $\int c_b\,dx$ is conserved to machine precision throughout the
simulation.

When the network becomes rigid, the domain is frozen at the transition height and
drying continues as flow and evaporation in a rigid porous layer. The
drying side of this stage extends the model of Joo et
al.~\cite{joo.etal2026StatedependentTransient}: the liquid
saturation $S(x,t)$ obeys a Darcy--capillary equation with the liquid
velocity
\begin{equation}
u_l = -\frac{\kappa_{\mathrm{rl}}(S)\,\kappa}{\mu_l}\frac{\partial p_l}{\partial x},
\qquad p_l = p_g - p_c(S),
\label{eq:model-darcy}
\end{equation}
where $u_l$ is the liquid velocity, $\kappa_{\mathrm{rl}}(S)$ is the
relative permeability of the liquid, $\kappa$ is the intrinsic
permeability of the porous film, $\mu_l$ is the liquid viscosity, and
$p_l$, $p_g$ and $p_c$ are the liquid, gas and capillary pressures. The
capillary pressure is given by a Leverett function of saturation, the
relative permeability is
$\kappa_{\mathrm{rl}} = \big((S-S_{\mathrm{il}})/(1-S_{\mathrm{il}})\big)^3$,
where $S_{\mathrm{il}}$ is the irreducible liquid saturation below which
the liquid no longer flows, non-equilibrium evaporation is distributed
over the pore surface, and vapor, air and heat equations close the
system; the same prescribed evaporative flux $\dot m(t)$ drives the surface. In this
stage the binder equation takes $\phi_l = \varepsilon S$, the
binder-advecting flux $F_b = \varepsilon S\,u_l$, and a
connectivity-limited diffusivity
\begin{equation}
D_b = D_0\,\varepsilon^{3/2}\sqrt{\kappa_{\mathrm{rl}}(S)},
\label{eq:model-dbeff}
\end{equation}
where $D_0$ is the same prefactor as in consolidation. Eq.~\eqref{eq:model-dbeff}
reuses the connectivity criterion the drying model already contains. In
pore emptying the liquid velocity of Eq.~\eqref{eq:model-darcy} is
proportional to $\kappa_{\mathrm{rl}}(S)$, so convective binder transport
goes to zero as the liquid network loses its connectivity at low
saturation; Eq.~\eqref{eq:model-dbeff} makes the binder diffusivity
decrease with the same factor, instead of leaving diffusion finite at
saturations where the liquid no longer forms a connected path through
the film.

The stage transition itself passes three things across the boundary: the
time at which it occurs, the film height, and the binder and temperature profiles at the end of
consolidation, which become the initial conditions of pore emptying. Nothing
about the binder field is reset or assumed; the surface enrichment built
during consolidation is inherited as is. This inheritance is the mechanism
the single-stage models of Section~\ref{sec:intro} cannot represent.

\subsection{Nondimensionalization}
\label{sec:model-nondim}

Nondimensionalizing Eq.~\eqref{eq:model-binder} with the changing film
thickness $H(t)$ as the length scale, $\xi = x/H(t)$, so that $\xi = 0$ is
the current collector and $\xi = 1$ the surface in both stages, puts both
stages in the same conservative form,
\begin{equation}
\frac{\partial(\hat c H)}{\partial t} = -\frac{\partial G}{\partial \xi},
\qquad
G = -\frac{\mathcal D}{H}\frac{\partial \hat c}{\partial \xi} + u(\xi)\,\hat c,
\qquad G(0)=G(1)=0,
\label{eq:model-collapse}
\end{equation}
where $\hat c = c_b/c_{b0}$ is the binder concentration normalized by its
initial uniform value $c_{b0}$, $G$ is the binder flux in the scaled
coordinate, $\mathcal D = D_0\phi_l^{1/2}$ is the diffusivity scale (the
prefactor $D_0$ times the liquid-fraction factor that remains after the
change of coordinate), and $u(\xi)$ is the velocity at which binder moves
relative to the scaled grid. The conditions $G(0) = G(1) = 0$ state that no
binder crosses the current collector or leaves the film.

The velocity $u(\xi)$ has two parts, and both matter for what follows. The
first is the physical advection velocity $F_b/\phi_l$: the liquid squeezed
toward the surface carries the binder with it. The second, $-\dot H\,\xi$,
is not a motion of material but an effect of the coordinate. As the film
thins, a material point that stays at the same physical height $x$ moves
to a larger scaled height $\xi = x/H$, so in the scaled coordinate it
appears to drift toward the surface at the rate $-\dot H\,\xi = \gamma\xi$.
This apparent drift is how the receding surface overtakes the binder: the
surface sweeps down over material points that cannot follow it, and binder
that cannot diffuse away fast enough accumulates beneath it. In the base
model both parts are proportional to $\xi$, zero at the collector and
largest at the surface, so that $u(\xi)/\gamma = (\phi_s/\phi_l + 1)\,\xi$
during consolidation; the pore-emptying velocity has the same linear form
to leading order. The drying history itself, $H/H_0$ during consolidation
and $S(t)$ during pore emptying, does not depend on thickness once time is
measured in units of the drying time $H/\gamma$, because the prescribed evaporative flux removes solvent at a rate set by $\dot m$ alone. With length,
velocity and time scaled in this way, a single dimensionless group
remains:
\begin{equation}
\ln\!\big(c_b^{\,\mathrm{surf}}/c_b^{\,\mathrm{int}}\big) = F(\mathrm{Pe}),
\qquad \mathrm{Pe} = \frac{\gamma H}{\mathcal D},
\label{eq:model-pe}
\end{equation}
where $c_b^{\,\mathrm{surf}}$ and $c_b^{\,\mathrm{int}}$ are the binder
concentrations at the surface and at the current-collector interface at
the end of drying, $\mathrm{Pe}$ is the P\'eclet number, the ratio of the
advective transport rate $\gamma$ to the diffusive rate $\mathcal D/H$
across the film, and $F$ is a function that the model computes but the
scaling argument does not specify. The left-hand side of
Eq.~\eqref{eq:model-pe}, the logarithm of the surface-to-interface
concentration ratio, is called the surface enrichment; a ratio of 3 gives
about 1.1 and a ratio of 6 about 1.8.

Because $\phi_l$ evolves during drying, quoting one $\mathrm{Pe}$ per
drying condition requires a reference state. Throughout this paper $H$ is
the dry film thickness and $\mathcal D$ is evaluated at the liquid fraction
of the stage transition, $\phi_l = 0.56$, so that each condition
corresponds to a single value of $\mathrm{Pe}$; any other consistent choice
rescales $\mathrm{Pe}$ by a constant factor and changes no comparison.
That the surface enrichment depends on drying rate and thickness only
through $\mathrm{Pe}$ was verified numerically: holding $\dot m H$ fixed
while changing the thickness by a factor of four leaves the predicted
surface enrichment identical to five decimal places. A P\'eclet-number dependence of the
surface enrichment during film consolidation was derived by Font et
al.~\cite{font.etal2018BinderMigration}; Eq.~\eqref{eq:model-pe} states
that the same single group also governs the enrichment left after pore
emptying, for the whole chain. This has two consequences for everything
that follows. First, at fixed material properties and microstructure, the
drying-rate series and the thickness range of Section~\ref{sec:intro} are
two ways of varying the same quantity, $\mathrm{Pe}$; between them they
cover $\mathrm{Pe}$ from 0.19 to 6.5, a factor of 34, and a model that
reproduces one of the two data sets must reproduce the other with the same
function $F$, because both are values of $F(\mathrm{Pe})$. Second,
$\mathrm{Pe}$ has a useful length reading: $\delta = \mathcal D/\gamma$ is
the thickness of the accumulation layer that forms where surface sweep-up
and back-diffusion balance, a length set by material properties and drying
rate alone, so $\mathrm{Pe} = H/\delta$ counts how many accumulation layers
fit in the film. The 70~\textmu m films have $\mathrm{Pe} = 0.19$--$1.06$
(the layer covers the film; diffusion-dominated), the 400~\textmu m films
at 2.2 and 6.5 (a surface boundary layer thinner than the film): the same
material and model, in qualitatively different regimes. With the drying
time $H/\gamma$ and the diffusion time $H^2/\mathcal D$, $\mathrm{Pe}$ is
the reciprocal of the Fourier number used in parts of the drying
literature.

\begin{figure}
  \centering
  \includegraphics[width=\linewidth]{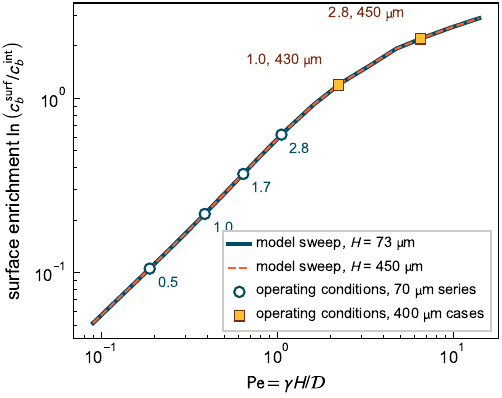}
  \caption{Predicted surface enrichment
$\ln(c_b^{\,\mathrm{surf}}/c_b^{\,\mathrm{int}})$ as a single function
$F(\mathrm{Pe})$ of the P\'eclet number $\mathrm{Pe} = \gamma H/\mathcal{D}$;
markers are model evaluations at the six measured operating conditions of
Section~\ref{sec:data}, not measurements.}
  \label{fig:model-collapse}
\end{figure}

Fig.~\ref{fig:model-collapse} shows this single-curve dependence over the
measured range: the drying rate swept at dry thicknesses of 73 and
450~\textmu m produces one curve $F(\mathrm{Pe})$, the two sweeps agreeing
to four decimal places, and the model evaluations at the six measured
operating conditions of Section~\ref{sec:data} lie on that curve between
$\mathrm{Pe} = 0.19$ and 6.5. In the figure the 73~\textmu m sweep is drawn
solid and the 450~\textmu m sweep dashed; open circles mark the
drying-rate series at 70~\textmu m, squares the 400~\textmu m cases, and
the point labels give $\dot m$ in g\,m$^{-2}$\,s$^{-1}$ and, for the
400~\textmu m cases, the dry thickness. The curve is computed with the
complete model, including the corrective terms of
\S\ref{sec:model-corrective}, and reads the simulated surface value
directly; the calendered surface reading of \S\ref{sec:model-obs}, used for
all comparisons in Section~\ref{sec:results}, lowers the two 400~\textmu m
values from 2.19 and 1.19 to 1.82 and 1.15.

\subsection{Corrective physics terms}
\label{sec:model-corrective}

\begin{figure}
  \centering
  \includegraphics[width=\linewidth]{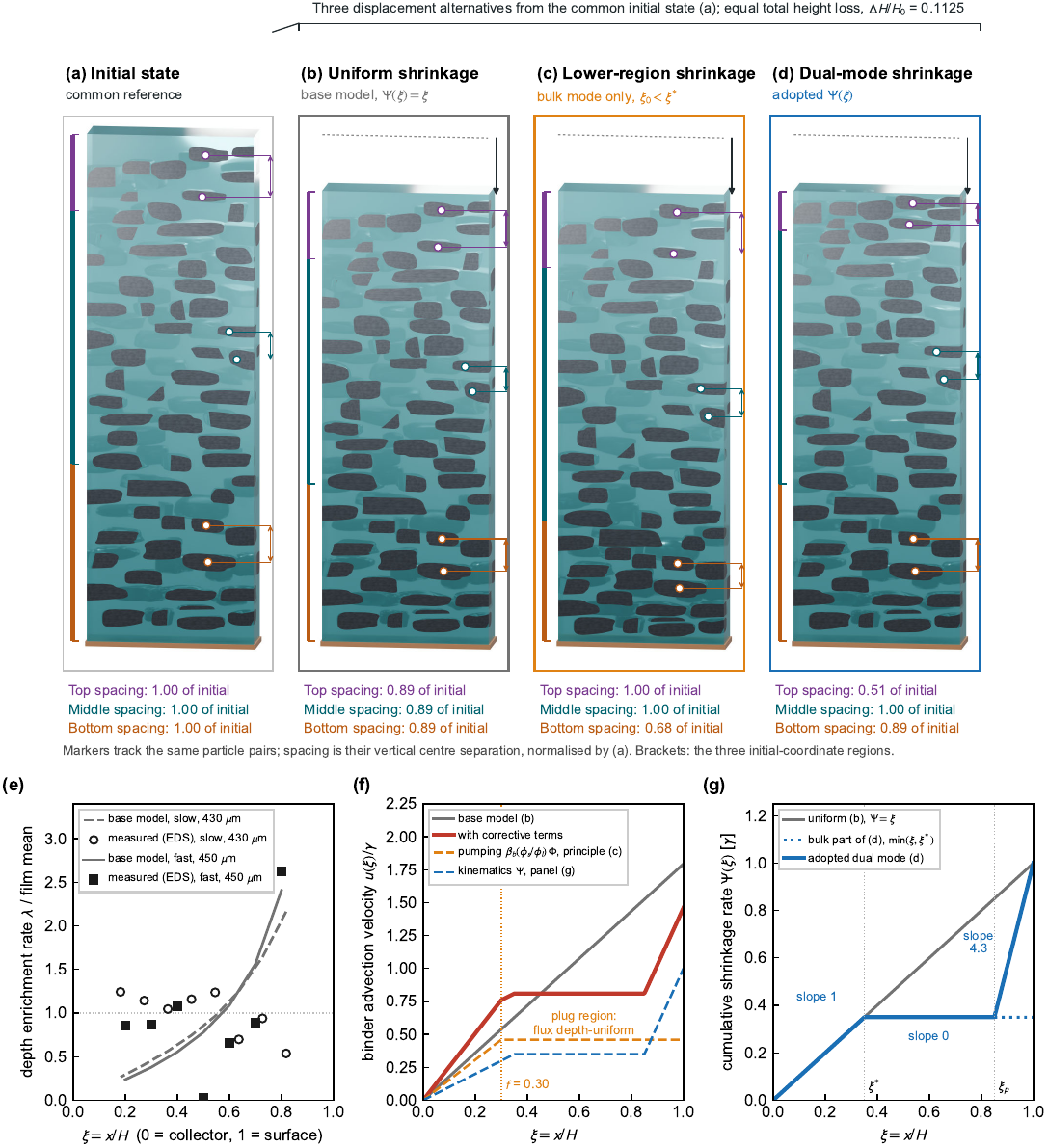}
  \caption{Shrinkage kinematics of film consolidation: (a) common initial
state; (b) uniform shrinkage of the base model; (c) shrinkage confined to
the lower region; (d) adopted dual-mode shrinkage; (e) local depth enrichment rate of the measured
binder profiles of the two thick films (430 and 450~\textmu m) and of the
base model, each divided by its film mean; (f) binder advection velocity
and (g) cumulative shrinkage rate for the same alternatives.}
  \label{fig:model-kinematics}
\end{figure}

The base chain above, with the linear-in-$\xi$ convection and unmodified
transport coefficients, does not reproduce the measurements in two
respects. First, in the cross-sections of the thickness series, the two
films dried to 430--450~\textmu m at 1.0 and
2.8~g\,m$^{-2}$\,s$^{-1}$~\cite{muller.etal2017InvestigationBinder}, the
EDS profiles show a nearly uniform depth enrichment rate, the relative
increase of binder concentration per unit depth
(\S\ref{sec:model-corr-pumping}), through the film, whereas the local
enrichment rate of the base chain falls by a factor of 8--10 from the
surface-side block to the deepest block (Fig.~\ref{fig:model-kinematics}e). Second, in the base chain binder transport stops, through
the connectivity factor of Eq.~\eqref{eq:model-dbeff}, at the same
effective saturation in every drying case, whereas the fast-dried
400~\textmu m film can be reproduced only if its transport stops almost at
the start of pore emptying, near $S \approx 1$
(\S\ref{sec:model-corr-threshold} and \S\ref{sec:results-ablation}). Five closed-form corrective terms are
therefore added.
All five act on the binder alone, on the advecting flux $F_b$ or on the
binder mobility; none modifies the liquid, solid or energy balances of the
drying chain, so they are effective descriptions of unresolved
microstructural physics, not additional conservation statements.
Fig.~\ref{fig:model-kinematics} pairs each consolidation term with the
function it introduces and with a three-dimensional picture of the
mechanism it stands for, and the pore-scale terms are drawn in the same
way where they are introduced. These pictures, like the sections of
Fig.~\ref{fig:model-overview}, are conceptual geometries drawn to explain
the terms: they are not reconstructions of measured microstructures, and
no computed saturation or concentration field is shown in them.
Sections~\ref{sec:model-corr-pumping}--\ref{sec:model-corr-immobilization}
introduce each term together with the specific observation that motivated
it, and \S\ref{sec:model-corr-roles} explains why two of the terms stop
binder motion. Each term reduces identically to the base model when
switched off, and the measurement that set the coefficients of each term
is listed in Table~\ref{tab:data-ledger} (\S\ref{sec:data-ledger}).

\subsubsection{Localized pumping profile}
\label{sec:model-corr-pumping}

The localized pumping profile reshapes the consolidation convection from
linear to plug. Together with the carrier coupling and the shrinkage
kinematics of the next two subsections, it gives the binder-advecting flux
\begin{equation}
F_b = \gamma\left[\,\beta_b\,\phi_s\,\Phi(\xi) + \phi_l\big(\Psi(\xi)-\xi\big)\right],
\qquad
\Phi(\xi) = \frac{\min(\xi, f)}{f}, \quad f = 0.3,
\label{eq:model-phi}
\end{equation}
where $\Phi(\xi)$ is the localized pumping profile; $f$ is the height of
the squeezing region as a fraction of the film thickness, so that liquid
is squeezed out of the particle network only for $0 \le \xi \le f$;
$\beta_b$ is the carrier coupling factor (\S\ref{sec:model-corr-coupling});
and $\Psi(\xi)$ is the shrinkage kinematics
(\S\ref{sec:model-corr-shrinkage}). The first term in the brackets is the
liquid pumped upward by consolidation together with the binder it carries;
the second is the change of flux caused by a non-uniform distribution of
the shrinkage, and it vanishes under uniform shrinkage, $\Psi = \xi$. With
$\beta_b = 1$ and $\Phi = \Psi = \xi$, Eq.~\eqref{eq:model-phi} returns the
base-model flux $F_b = \gamma\phi_s\xi$. This subsection concerns $\Phi$.

The shape of $\Phi$ follows from liquid continuity: the upward liquid flux
at any height equals the volume of liquid squeezed out per unit time below
that height. If the network shrinks uniformly, liquid is squeezed out at
every height and the flux grows linearly with $\xi$. If consolidation is
confined to the lower part of the film, $\xi \le f$, all liquid is
expelled there and passes through the region above without further
addition, so above $f$ the flux is the same at every height, a plug flow.
Fig.~\ref{fig:model-kinematics}c shows this limiting case with the
particles drawn as rigid spheres: the film loses height only in the lower
region, where the particles move closer together, while the particles
above all move down together and keep their spacing. $\Phi$ takes from
this limiting case only the shape of the pumped liquid flux; how the loss
of height is shared between heights is described separately by the
shrinkage kinematics $\Psi$ of \S\ref{sec:model-corr-shrinkage}.

The motivating observation is the measured depth enrichment rate
$\lambda = u/\mathcal D$, the local relative increase of binder
concentration per unit depth, which in the quasi-steady balance between advection
and back-diffusion is proportional to the local binder advection velocity
$u$. In the 400~\textmu m cross-sections $\lambda$ is nearly
depth-uniform: the enrichment-rate ratio, the mean of
$\lambda$ over the surface-side half of the film divided by its mean over
the deep-side half, is 0.74 and 1.12 in the slow- and fast-dried films,
whereas the base model gives 3.0 and 3.8
(\S\ref{sec:results-endpoint}). Fig.~\ref{fig:model-kinematics}e shows the shapes of these profiles,
each divided by its film mean: the measured values scatter without a trend
in height, while those of the base model rise steadily from about 0.25
at the deepest block to 2.2--2.4 at the surface-side block. A depth-uniform $\lambda$ therefore
indicates that the advection velocity is plug-like above the pumping
region. The term is named by what it does, localized pumping, and no
microstructural picture of where consolidation concentrates is claimed;
the same reshaping is applied to the pore-emptying velocity profile, which
is also linear in $\xi$ before reshaping.

\subsubsection{Carrier coupling}
\label{sec:model-corr-coupling}

\begin{figure}
  \centering
  \includegraphics[width=\linewidth]{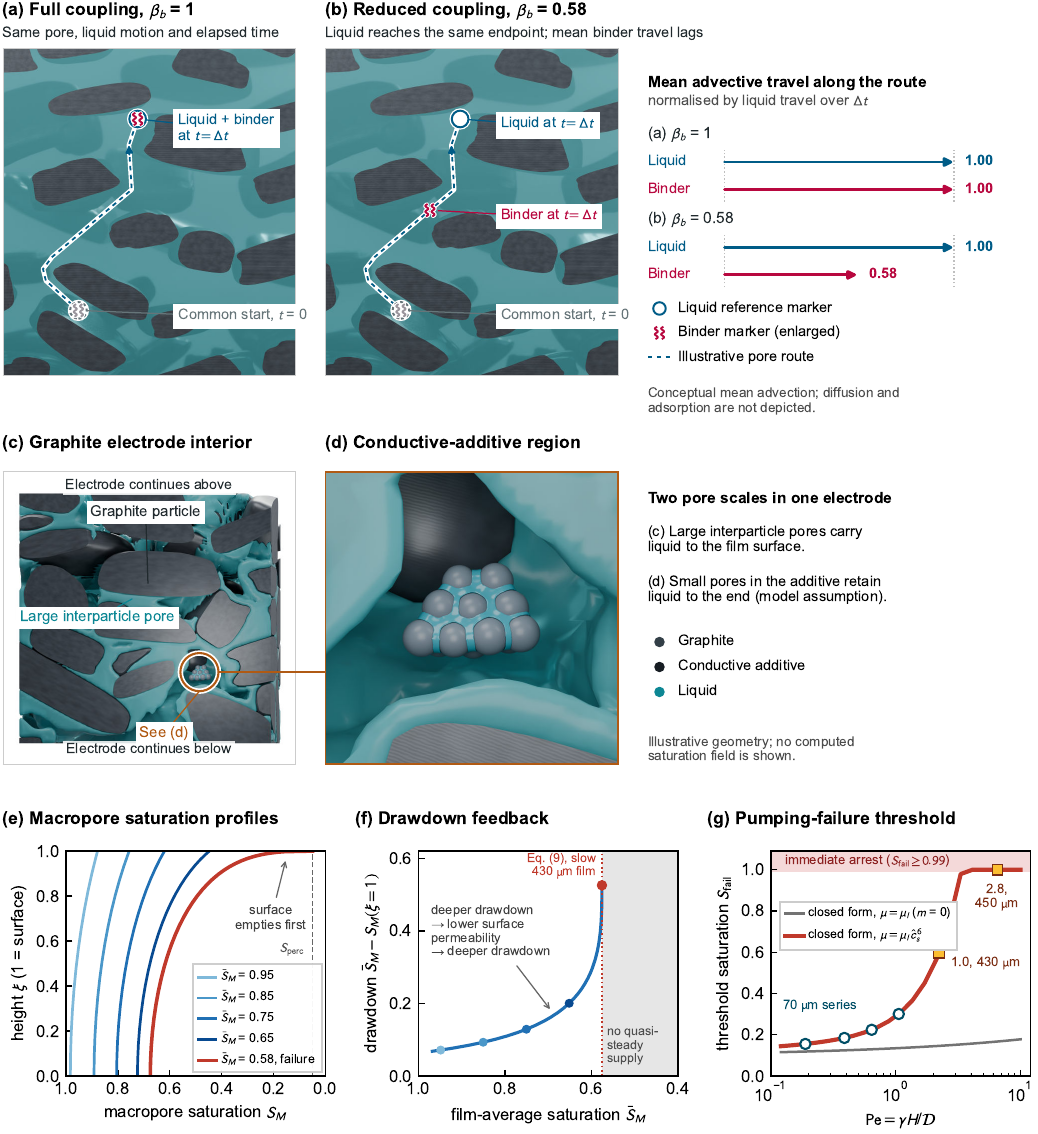}
  \caption{Pore-scale transport and retention: (a, b) carrier coupling in
one internal pore at $\beta_b$ = 1 and 0.58; (c, d) interparticle pores and
the conductive-additive region of the same electrode; (e) macropore
saturation profiles at decreasing film-average macropore saturation
$\bar S_M$ and (f) the surface drawdown against $\bar S_M$, an
illustration of the closure of Eq.~\eqref{eq:model-sfail} and not a
computed drying field; (g) pumping-failure threshold against
$\mathrm{Pe}$ with the six measured conditions.}
  \label{fig:model-pore}
\end{figure}

The binder is advected at a reduced velocity
relative to the liquid, $u_b = \beta_b\,u_l$ with $\beta_b = 0.58$, where
$u_b$ is the binder velocity; the factor is applied identically in both
stages and appears in Eq.~\eqref{eq:model-phi} as the prefactor of the
pumping term. In Fig.~\ref{fig:model-kinematics}f, discussed at the end of
\S\ref{sec:model-corr-shrinkage}, it multiplies the pumping curve by the
same factor at every height, so it changes the magnitude of that curve and
leaves its shape unchanged. Fig.~\ref{fig:model-pore}a,b
illustrates the factor in one pore: over the same elapsed time the mean
binder travel along the liquid path is $\beta_b$ times the liquid travel.
The factor scales the
binder-advecting flux $F_b$ only; the liquid flux and the solvent mass
balance of the drying chain are unchanged, so no statement about liquid
conservation is affected by $\beta_b \neq 1$. The value is a fitted
parameter, determined together with $f$, $\xi^{*}$ and $\xi_p$ from the
400~\textmu m endpoint profiles (Table~\ref{tab:data-ledger}). The
motivating observation is the interface binder concentration of the
fast-dried 400~\textmu m film: lowering $\beta_b$ reduces the upward
transport of binder and raises that concentration, which at
$\beta_b = 0.58$ reaches 0.687, at the lower edge of its measured band
0.69--0.78 (Table~\ref{tab:results-ablation}).

\subsubsection{Dual-mode shrinkage}
\label{sec:model-corr-shrinkage}

In the base model the film shrinks uniformly, and the shrinkage kinematics
of Eq.~\eqref{eq:model-phi} is then simply $\Psi(\xi) = \xi$. The dual-mode
shrinkage replaces it by
\begin{equation}
\Psi(\xi) = \min(\xi, \xi^{*})
+ \big(1-\xi^{*}\big)\,
  \max\!\Big(0, \min\!\Big(1, \frac{\xi-\xi_p}{1-\xi_p}\Big)\Big),
\quad \xi^{*} = 0.35,\ \xi_p = 0.85,
\label{eq:model-psi}
\end{equation}
where $\xi^{*}$ is the fraction of the total shrinkage taken up by the
lower film and $\xi_p$ is the scaled height at which the surface
deposition layer begins. $\Psi(\xi)$ is the cumulative shrinkage rate of
the film column below $\xi$, in units of the surface recession rate
$\gamma$, so its gradient $\partial\Psi/\partial\xi$ is the local
consolidation rate: where the gradient is 1 the network contracts as under
uniform shrinkage, where it is 0 material moves as a block without
particles approaching one another, and $\Psi(1) = 1$ states that the film
as a whole must shrink at the rate $\gamma$ at which solvent leaves,
whatever the distribution. Under uniform shrinkage every material point
approaches every other like marks on a contracting elastic band.
Fig.~\ref{fig:model-kinematics}g shows how Eq.~\eqref{eq:model-psi}
instead concentrates consolidation in the lower film (bulk mode,
$\xi < \xi^{*}$, gradient 1), lets the middle of the film move as a block
(gradient 0), and places the remaining fraction $1-\xi^{*}$ of the
shrinkage in a thin packing layer at the surface (deposition ramp above
$\xi_p$, gradient 4.3). Panels (b)--(d) of the same figure show the three
alternatives on one rigid-particle packing with the same total height
loss: under uniform shrinkage every tracked pair of particles approaches
by the same fraction; with the loss confined to the lower region only the
bottom pair approaches; under the adopted dual mode the bottom and top
pairs approach while the middle pair keeps its spacing. Note that
$\Psi$ is cumulative shrinkage
kinematics and not a cumulative binder amount. The motivating observations
are two features of the 400~\textmu m cross-sections that the uniform
kinematics cannot reproduce together: the sharp rise of the binder
concentration in the topmost block of each measured profile, the surface
spike, and the binder concentration at the interface compared on the
absolute, mass-conserving scale of \S\ref{sec:results-endpoint} rather
than as a normalized shape. The block-like middle keeps the interior
enrichment rate flat, while the deposition layer, where the liquid squeezed out at the
surface evaporates and leaves its binder behind, restores the measured
surface level. The packing layer is an effective description of how the
shrinkage is distributed, not an observed microstructure.

Fig.~\ref{fig:model-kinematics}f shows the binder advection velocity $u$ at
the end of consolidation that results from the three terms of
Eq.~\eqref{eq:model-phi}, against the base-model line. Dividing
Eq.~\eqref{eq:model-phi} by $\phi_l$ and adding the coordinate drift
$\gamma\xi$ of \S\ref{sec:model-nondim} gives
$u/\gamma = \beta_b(\phi_s/\phi_l)\Phi + \Psi$, so the corrected profile is
the sum of the pumping part $\beta_b(\phi_s/\phi_l)\Phi$
(\S\ref{sec:model-corr-pumping} and \S\ref{sec:model-corr-coupling}) and
the kinematic part $\Psi$ (this subsection); in the base model the same sum
is $(\phi_s/\phi_l)\xi + \xi$.

\subsubsection{Dual-porosity pumping-failure threshold}
\label{sec:model-corr-threshold}

In pore emptying, binder
transport terminates as an event: when the domain-average saturation falls
below a case-dependent threshold $S_{\mathrm{fail}}$, both $F_b$ and $D_b$ for the
binder are switched off and the profile freezes in place. The threshold is
computed in closed form from a dual-porosity picture of the electrode pore
space: a macropore network between the graphite particles, through which
liquid is pumped over long distances, and a nanopore reservoir in the
conductive additive, of pore-volume fraction $w_n$, whose capillary
pressure is high enough that it remains saturated to the end.
Fig.~\ref{fig:model-pore}c,d shows the two scales inside one electrode:
the large pores between graphite particles that supply liquid to the
surface, and the small interstices of a conductive-additive aggregate that
hold liquid back. The total
saturation $S$ and the macropore saturation $S_M$ are related throughout by
$S = w_n + (1-w_n)\,S_M$. The macropore network is given a percolation
form of relative permeability,
$\kappa_{\mathrm{rl},M} = \big((S_M - S_{\mathrm{perc}})/(1 - S_{\mathrm{perc}})\big)^{p}$,
where $S_{\mathrm{perc}}$ is the macropore saturation at which the liquid
network loses its long-range connectivity and $p$ is the percolation
exponent. Supplying the evaporative demand $\gamma$ at the surface by
Darcy flow requires a capillary-pressure gradient, hence a saturation
gradient: liquid must flow from the wetter interior to the surface, and
the surface saturation lies below the film average by a drawdown
$\Delta S_M \approx B/\kappa_{\mathrm{rl},M}$ (Fig.~\ref{fig:model-pore}e), where $B$ is the
ratio of the pressure drop needed to pump liquid across the film to the
capillary pressure available to drive it. As drying lowers the average
saturation the drawdown deepens, the surface permeability falls, and the
drawdown needed to sustain the same demand grows further; this feedback
runs away (Fig.~\ref{fig:model-pore}f), and the surface reaches $S_{\mathrm{perc}}$ at a finite average
saturation. Balancing the quasi-steady drawdown against the percolation
closure gives that failure point as follows:
\begin{equation}
S_{M,\mathrm{fail}} = S_{\mathrm{perc}} + \Big[\chi\,B\,(1-S_{\mathrm{perc}})^{p}\Big]^{\frac{1}{p+1}},
\qquad
B = \frac{\gamma\,\mu(\hat c_s)\,H}{\kappa\,p_{c0}\,J},
\qquad
\mu(\hat c_s) = \mu_l\,\hat c_s^{\,m},
\label{eq:model-sfail}
\end{equation}
where $S_{M,\mathrm{fail}}$ is the film-average macropore saturation at
failure, $\chi$ is a prefactor of order unity that absorbs the geometric
uncertainty of the closure, $\mu(\hat c_s)$ is the liquid viscosity at the
surface binder concentration $\hat c_s$ reached at the end of
consolidation, $m$ is the concentration--viscosity exponent,
$p_{c0} = \sigma\sqrt{\varepsilon/\kappa}$ is the capillary pressure scale
with $\sigma$ the surface tension, and $J$ is the Leverett function
evaluated at intermediate saturation. The threshold in total saturation is
$S_{\mathrm{fail}} = w_n + (1-w_n)\,\min(S_{M,\mathrm{fail}},1)$. The values
used are $w_n = 0.05$, $S_{\mathrm{perc}} = 0.05$, $p = 2$, $\chi = 3$,
$m = 6$ and $J = 0.36$. Fig.~\ref{fig:model-pore}e,f shows this closure
for a simplified quasi-steady supply, in which the
liquid flux through each height is proportional to the height and $B$ is
chosen so that failure occurs at the $S_{M,\mathrm{fail}}$ of the slowly
dried 430~\textmu m film: the macropores empty from the surface
downward, and the drawdown grows slowly at first and then steeply as the
average approaches that value. Fig.~\ref{fig:model-pore}g shows the
resulting threshold against $\mathrm{Pe}$. With a constant viscosity
($m = 0$, gray curve) the threshold varies little across the measured
range; the concentration-dependent viscosity, evaluated at the surface
concentration that consolidation has built, is what makes it
case-dependent (red curve). Fast-dried thick films, the high-$\mathrm{Pe}$
cases at the right of Fig.~\ref{fig:model-pore}g, arrive at pore emptying
with a concentrated, highly viscous surface layer, so their carrier
network fails immediately ($S_{\mathrm{fail}} \approx 1$ for the fast-dried
400~\textmu m case, shaded region), while slowly dried thin films, the
low-$\mathrm{Pe}$ cases at the left, have dilute surfaces and keep pumping
to deep saturations ($S_{\mathrm{fail}} = 0.16$--$0.30$ for the
70~\textmu m series; 0.60 for the slow 400~\textmu m case). Two
questions arise here. First, does $\kappa_{\mathrm{rl}}$ not already stop
transport at low saturation? It does, but at the same effective saturation
for every case; the arrest saturations of the drying cases range from
$S \approx 1$ down to $S \approx 0.16$, and that spectrum is what the
threshold supplies.
Second, is dual porosity in the governing equations? No: it acts only as
this threshold generator. The drying equations, $\kappa_{\mathrm{rl}}$ and
$S_{\mathrm{il}}$ are unmodified; the closed form computes a single number
per case, at which binder transport is switched off. The prefactor $\chi$
absorbs the O(1) geometric uncertainty of the closure, including the
replacement of the local Leverett slope by the constant $J$; that its
value is of order unity supports the plausibility of the closure, but the
closure is not a first-principles result.

The surface concentration $\hat c_s$ in the viscosity law is read on the
total-volume basis of Eq.~\eqref{eq:model-binder}. The binder
concentration in the liquid itself, which physically sets the viscosity,
is larger by the factor $\phi_{l0}/\phi_l$ at the transition, where
$\phi_{l0}$ is the initial liquid fraction. Because the scaled
consolidation trajectory is the same in every case
(\S\ref{sec:model-nondim}), this factor is a constant: it rescales $B$ by
the same amount in every case and is absorbed in the prefactor $\chi$.

\subsubsection{Binder immobilization function}
\label{sec:model-corr-immobilization}

\begin{figure}
  \centering
  \includegraphics[width=\linewidth]{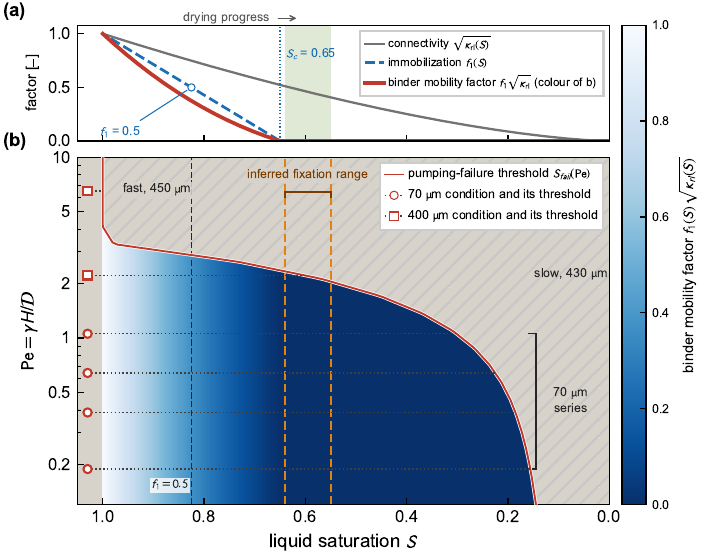}
  \caption{(a) Immobilization function $f_1(S)$, connectivity factor
$\sqrt{\kappa_{\mathrm{rl}}(S)}$ and their product, the binder mobility
factor; (b) binder arrest map: distribution of the binder mobility factor
on the saturation--P\'eclet plane.}
  \label{fig:model-arrestmap}
\end{figure}

The last term acts not on the
carrier liquid but on the binder itself. All binder mobility in pore
emptying, convective velocity and diffusivity alike, is multiplied by a
saturation-dependent factor
\begin{equation}
f_1(S) = \left(\frac{S - S_c}{1 - S_c}\right)^{q}
\ \ \text{for } S > S_c, \qquad f_1 = 0 \ \text{otherwise},
\qquad S_c = 0.65,\ q = 1,
\label{eq:model-f1}
\end{equation}
where $S_c$ is the arrest point, the saturation at which the binder
mobility reaches zero, and $q$ is the immobilization exponent; the drying
physics is untouched. The form is that of the power law used by Coussot
\cite{coussot2000ScalingApproach} for the permeability of an unsaturated
porous medium, which vanishes below a critical saturation.
Fig.~\ref{fig:model-arrestmap}a
shows $f_1$ together with the connectivity factor
$\sqrt{\kappa_{\mathrm{rl}}}$ of Eq.~\eqref{eq:model-dbeff} and their
product, the binder mobility factor $f_1\sqrt{\kappa_{\mathrm{rl}}}$, which
multiplies all binder transport in pore emptying and is also the colour of
the map in panel (b). With $S_c = 0.65$, $f_1$ falls to one half at
$S = 0.825$, the binder is effectively arrested below $S \approx 0.7$, and
$f_1$ reaches zero at $S_c = 0.65$. The inferred fixation range, $S$ =
0.55--0.64, is the saturation range by which the binder is inferred to have
become fixed: in the switch-time experiments of both sources, a change of
drying rate after this point no longer changed the interface. The range is
obtained by converting the characteristic solvent volume fractions reported
in \cite{jaiser.etal2016InvestigationFilm,jaiser.etal2017DevelopmentThreestage}
to saturation (\S\ref{sec:data-sets}, \S\ref{sec:results-arrest}), and it lies
just below $S_c$, where the binder mobility factor has fallen to zero.
Unlike every other term
in this section, $f_1$ is not constrained by the endpoint measurements at
all: \S\ref{sec:results-ablation} shows that removing it leaves all
endpoint predictions essentially intact and destroys only the response to
the switching time. It is therefore the one term calibrated on the
switch-time experiment: its arrest point $S_c$ is fitted to the midpoint of
the switch-time response~\cite{jaiser.etal2016InvestigationFilm}
(Table~\ref{tab:data-ledger}), whereas the three-stage schedule
experiment~\cite{jaiser.etal2017DevelopmentThreestage} of
Section~\ref{sec:results-hlh} is not used for this or any other calibration.

\subsubsection{Two arrest mechanisms}
\label{sec:model-corr-roles}

Two of the five terms stop binder motion, and because they explain
different observations, neither can stand in for the other. The threshold
of Eq.~\eqref{eq:model-sfail} concerns the carrier, the liquid network that
carries the binder: it marks the percolation event at which this network
fails hydraulically, and its position depends on how concentrated the
surface became during consolidation and hence on the drying case. The
immobilization function of Eq.~\eqref{eq:model-f1} concerns the binder
itself: the binder drops out of transport as saturation falls, a material
property with one value of $S_c$ for all cases. In the fast-dried thick
film the carrier network fails at $S \approx 1$, where $f_1 \approx 1$, so
the immobilization function alone would let the binder keep moving. In the
thin films the carrier network remains connected down to
$S = 0.16$--$0.30$, far below the inferred fixation range, so the threshold
alone would let the binder move for too long.
Section~\ref{sec:results-ablation} quantifies this separation of roles by
removing the two mechanisms one at a time.

Fig.~\ref{fig:model-arrestmap}b shows that in most of the six measured
conditions it is the immobilization function that stops the binder, and
that the threshold acts first only in the fast-dried 450~\textmu m film.
The horizontal axis is the saturation shared with panel (a), with drying
progressing to the right. The vertical axis is $\mathrm{Pe}$, and the
background colour is the binder mobility factor
$f_1(S)\sqrt{\kappa_{\mathrm{rl}}(S)}$. The gray region is where the
threshold $S_{\mathrm{fail}}(\mathrm{Pe})$ of Eq.~\eqref{eq:model-sfail}
has stopped binder transport; the black dashed line marks $S = 0.825$,
where $f_1 = 1/2$; the two orange dashed lines bound the inferred fixation
range $S$ = 0.55--0.64 (shaded in panel a); and the markers place the six
measured drying conditions. The background colour fades to zero along the
same saturation at every $\mathrm{Pe}$, because the immobilization function
and the connectivity factor are both properties of the material alone. The
boundary of the gray region, by contrast, moves toward higher saturation as
$\mathrm{Pe}$ increases, from deep saturations in the slowly dried thin
films to $S \approx 1$ in the fast-dried thick film. In the 70~\textmu m
series (0.16--0.30) and the slowly dried 430~\textmu m film (0.60) the
threshold lies below $S = 0.65$, where the immobilization function reduces
the mobility to zero, so in these five conditions the saturation at which
the binder stops is set by the immobilization function, independently of
$\mathrm{Pe}$. Only in the fast-dried 450~\textmu m film does the threshold
come first, at $S \approx 1$, stopping the binder as soon as pore emptying
begins.

\subsection{Observation operators}
\label{sec:model-obs}

What the model produces is the distribution of dissolved binder in the film
at the moment binder transport stops; what was measured is the film after
it had dried completely and been calendered, read by EDS. Comparing the two
requires accounting for three things that happen to the sample in between:
the remaining solvent evaporates and leaves the binder behind as a solid,
calendering compresses and mixes the layer near the surface, and the
surface and the interface are measured on different faces. All three are
handled only at comparison time and do not enter the physics of the
simulation.

The first is a question of what is compared with what. EDS reads the solid
PVDF distribution of the completely dried film, whereas the model variable
is the binder in solution. Once binder transport has stopped, the remaining
solvent evaporates without moving the binder further, so the dissolved
binder is taken to precipitate in place, and the $c_b$ field at the moment
transport stops is compared directly with the measured solid distribution.

Second, all cross-sectioned samples of
\cite{muller.etal2017InvestigationBinder} were calendered before
measurement. Calendering presses the film by several tens of micrometres
in the thickness direction, compressing and mixing the layer near the
surface, so the binder concentration measured in the top block is close to
the average of the mixed layer rather than to the surface concentration
right after drying. The surface reading of the model is therefore taken as
the conservative average of the simulated profile over the layer down to a
depth $d$ below the surface:
\begin{equation}
\hat c_{\mathrm{surf}}^{\,\mathrm{obs}}
= \frac{1}{d}\int_{H-d}^{H} \hat c\,dx,
\qquad d = \alpha\,\Delta H_{\mathrm{cal}},
\label{eq:model-calender}
\end{equation}
where $\Delta H_{\mathrm{cal}}$ is the measured calendering compression of
each film (86 and 46~\textmu m for the fast- and slow-dried 400~\textmu m
films; for the 70~\textmu m series converted from the reported porosity
reduction), and $\alpha$ is the partial homogenization coefficient, the
ratio of the mixing depth to the compression. With $\alpha = 0$, the depth
$d$ in Eq.~\eqref{eq:model-calender} is zero and the surface reading is not
the average of a mixed layer but the value at the single top point of the
model profile, $x = H$: calendering then only presses the film without
mixing the near-surface layer, so the concentration at the top right after
drying is read as it is. Because the source does not report the
calendering conditions, $\alpha$ was set from the surface reading to 0.5,
that is, the layer down to half the compression depth is taken to be mixed
homogeneously. The mixing depth is then 43~\textmu m in the fast-dried film
and 23~\textmu m in the slow-dried film, within the top blocks that the
source excluded from its analysis (45 and 39~\textmu m). This calendering observation
operator changes no simulated field, only the surface reading. Switching
the operator off shows that what it corrects is the reading at one surface
point. With $\alpha = 0$ and $D_0$ re-fitted on the drying-rate series, the
RMS difference between model and measurement over all interior blocks of
the two 400~\textmu m cross-sections stays nearly the same, 0.093 against
0.091. Only the comparisons that depend on the surface reading degrade: the
change of the surface-to-interface ratio with drying rate between the two
400~\textmu m films, model divided by measurement, grows from 1.01 to 1.51.

Third, the surface and interface values of the drying-rate series are
measured on two different faces. The surface value is EDS on the top face
of the film, and the interface value is EDS on the bottom face exposed by
peeling off the copper current-collector foil. The source itself notes that
comparing the two values directly can be misleading. On the top face the
thin binder film covering the particles can be expected to dominate, and on
the peeled face the binder gathered at the contact points. The supplement
of the source contains measurements of two samples of nearly the same
composition (drying rates 0.6 and 4.1~g\,m$^{-2}$\,s$^{-1}$) by both XPS and
EDS, which show how much the measurement method changes the reading. On the
same sample, the surface-to-interface ratio measured by EDS is smaller than
that measured by XPS, 0.75--0.89 times its value; that is, EDS reads the
interface side as relatively richer in binder compared with the surface.
Because the two methods have different information depths (XPS the top few
nanometres, EDS the order of micrometres), their absolute concentrations
cannot be compared, and these data cannot tell whether the surface or the
interface reading is biased. When the surface enrichment of each
measurement is plotted against drying rate, the points form a straight
line. With the slope of this line, the increase of the surface enrichment
when the drying rate rises by 1~g\,m$^{-2}$\,s$^{-1}$, called the rate
sensitivity $k$ of the surface enrichment, $k$ is 0.203 for XPS, 0.252 for
EDS on the same samples and 0.212 for the EDS series used for calibration,
all within 0.20--0.25, whereas the intercepts, $-0.09$, $-0.41$ and
$-0.22$, differ by more than a factor of four. What the measurement method
changes is mainly the height of the line. This series is therefore compared
only through its rate sensitivity, and the intercept is excluded from
calibration.

\subsection{Numerics}
\label{sec:model-numerics}

The full chain was solved by a conservative finite-volume scheme on a
uniform $\xi$ grid of 500 cells, which preserves the binder inventory to
machine precision, with the drying trajectories of both stages supplied by
the model of \cite{joo.etal2026StatedependentTransient}. Halving or
doubling the cell count at the adopted parameter set changes every model
value of Table~\ref{tab:results-comparison} by less than 0.3\% and the
response midpoint by less than 0.1~s. The same
physics was ported to a COMSOL implementation used for the drying-side
solutions; the two agree to a relative parity of $2\times10^{-9}$ on the
adopted configuration, and the end-to-end chain reproduces the comparison
values computed by the screening solver within $\pm 3\%$.

%% file: sections/03_data_calibration.tex
\section{Data and calibration}
\label{sec:data}

\subsection{Datasets}
\label{sec:data-sets}

The model is compared with four experiments from three sources on the same
material system, graphite anodes with PVDF binder cast from NMP, and the
four experiments vary the three operating variables of the drying oven one
at a time (Table~\ref{tab:data-sets}). Drying intensity and loading are
covered by the two series of \cite{muller.etal2017InvestigationBinder}, and
scheduling by the two experiments of
\cite{jaiser.etal2016InvestigationFilm} and
\cite{jaiser.etal2017DevelopmentThreestage}. All experimental conditions
are public values from the source papers, and no private data is used.

\begin{table}
\caption{The four experiments compared with the model and their use}
\label{tab:data-sets}
\centering
\footnotesize
\setlength{\tabcolsep}{3pt}
\begin{tabular}{>{\raggedright\arraybackslash}p{2.0cm}>{\raggedright\arraybackslash}p{1.8cm}>{\raggedright\arraybackslash}p{3.6cm}>{\raggedright\arraybackslash}p{3.0cm}>{\raggedright\arraybackslash}p{2.2cm}}
\toprule
Source and experiment & Operating variable varied & Conditions & Measurement compared & Use \\
\midrule
\cite{muller.etal2017InvestigationBinder}, drying-rate series (A70) &
Drying rate &
4 rates, $\dot m$ = 0.5--2.8 g\,m$^{-2}$\,s$^{-1}$ (a factor of 5.6), dry
thickness 73--75~\textmu m &
Surface-to-interface binder ratio at each rate (EDS; rate sensitivity
only, \S\ref{sec:model-obs}) &
Calibration of $D_0$ \\
\addlinespace
\cite{muller.etal2017InvestigationBinder}, series A400 &
Thickness and loading &
2 rates (1.0 / 2.8 g\,m$^{-2}$\,s$^{-1}$), dry thickness 430 and
450~\textmu m (six times the A70 thickness) &
Binder concentration cross-sections by depth block; surface and
interface EDS values &
Calibration of the corrective-term coefficients \\
\addlinespace
\cite{jaiser.etal2016InvestigationFilm}, switch-time experiment &
Time at which the drying rate is lowered &
Isothermal 76.5~\textdegree C; start at 1.19 g\,m$^{-2}$\,s$^{-1}$, switch
to 0.52 g\,m$^{-2}$\,s$^{-1}$ at time $t_{\mathrm{trans}}$; dry film
70.5 g\,m$^{-2}$ ($\approx$78~\textmu m), porosity 0.59 &
Final interface binder and peel strength against switch time (source
Fig.~3) &
Calibration of $S_c$ on the response midpoint; drying times predicted \\
\addlinespace
\cite{jaiser.etal2017DevelopmentThreestage}, three-stage experiment &
Position of the low-rate stage &
Same material system and two rates; high--low--high with the low-rate
stage between liquid volume fractions $\phi_{t1}$ =
0.63/0.60/0.58/0.56 and $\phi_{t2}$ = 0.45; dry film 72 g\,m$^{-2}$ &
Adhesion (source Fig.~5); drying times calculated by the source (source
Table~4); characteristic solvent volume fractions &
Not used (predicted, \S\ref{sec:results-hlh}) \\
\bottomrule
\end{tabular}
\end{table}

The two series of \cite{muller.etal2017InvestigationBinder} measure where
the binder is located in the dried film. The drying-rate series (A70) dried
films about 70~\textmu m thick at four drying rates and measured the binder
concentration by EDS on the top face of the film and on the bottom face
exposed by peeling off the current-collector foil. The A400 series dried
films six times thicker, 430 and 450~\textmu m, at 1.0 and
2.8~g\,m$^{-2}$\,s$^{-1}$, respectively, divided each cross-section into
blocks of equal depth, and reported the relative binder concentration
measured in each block. The samples of both series were calendered before
measurement (\S\ref{sec:model-obs}).

The switch-time experiment of \cite{jaiser.etal2016InvestigationFilm}
measures how the interface binder responds to the time at which the drying
rate is changed during drying. Films about 78~\textmu m thick were dried
starting at 1.19~g\,m$^{-2}$\,s$^{-1}$, the rate was lowered to
0.52~g\,m$^{-2}$\,s$^{-1}$ at the switch time $t_{\mathrm{trans}}$, and the
final interface binder and peel strength were measured for a range of
switch times. The source reports the last switch time at which the adhesion
still equals that of a film dried slowly throughout, $t_{c,1} = 28$~s, and
the first at which it equals that of a film dried fast throughout,
$t_{c,2} = 44$~s.

The three-stage experiment of \cite{jaiser.etal2017DevelopmentThreestage}
varies where the low-rate stage is placed during drying. Films of the same
material system with a dry loading of 72~g\,m$^{-2}$ were dried at the high
rate, switched to the low rate when the liquid volume fraction reached
$\phi_{t1}$, and returned to the high rate at $\phi_{t2} = 0.45$. The source
defines this liquid volume fraction $\phi$ as the volume of liquid divided
by the volume of liquid and solids, so the air that enters emptied pores is
not counted. During film consolidation $\phi$ falls as the film shrinks, and
it equals the dry-film porosity $\varepsilon$ when the pores begin to empty;
during pore emptying it is related to the liquid saturation $S$ by
$S = \phi(1-\varepsilon)/[\varepsilon(1-\phi)]$, so that $\phi = 0.45$
corresponds to $S = 0.57$ at $\varepsilon = 0.59$. Adhesion was measured in
four scenarios with $\phi_{t1}$ = 0.63, 0.60, 0.58 and 0.56, and the drying
time of each scenario was calculated by the source from the measured drying
rates. The source also repeated the switch-time experiment at the same two
rates and converted its characteristic times to solvent volume fractions,
one of which it names the binder immobilization point
(\S\ref{sec:results-arrest}).

Film thickness, solids content and porosity enter as literature values and
are not fitted. The slurry composition reported in
\cite{jaiser.etal2016InvestigationFilm} coincides with the composition
parameters of the model, so no compositional conversion is needed to turn
the reported conditions into simulation inputs.

These experiments did not publish the air-side conditions, such as nozzle
velocity and humidity; what is published is the drying
rate itself. The model is therefore driven by a prescribed evaporative flux
$\dot m(t)$ (\S\ref{sec:model-governing}). This is the only way the
experiments can be reproduced, and the same variable connects the
process-level model to the electrode model (\S\ref{sec:discussion-limits}).

\subsection{Calibration summary}
\label{sec:data-ledger}

The parameters of the model were fitted to measurements on dried films and
to the midpoint of the switch-time response, whereas the drying times of
the two schedule experiments and the outcome of the three-stage experiment
were predicted without fitting. Table~\ref{tab:data-ledger} lists, for
each degree of freedom, the measurement that set its value.

\begin{table}
\caption{Measurement that set each degree of freedom}
\label{tab:data-ledger}
\centering
\footnotesize
\setlength{\tabcolsep}{4pt}
\begin{tabular}{>{\raggedright\arraybackslash}p{4.0cm}>{\raggedright\arraybackslash}p{3.0cm}>{\raggedright\arraybackslash}p{5.6cm}}
\toprule
Degree of freedom & Value & Measurement that set the value \\
\midrule
$D_0$ & $2.52\times10^{-10}$ m$^2$\,s$^{-1}$ & Rate sensitivity of the
drying-rate series \\
\addlinespace
$\beta_b$ (carrier coupling), $f$, $\xi^*$, $\xi_p$ & 0.58 / 0.3 / 0.35 /
0.85 & Cross-sections of the A400 series (end blocks and depth enrichment
rate) \\
\addlinespace
$w_n$, $S_{\mathrm{perc}}$, $p$, $\chi$, $m$ (threshold,
Eq.~\ref{eq:model-sfail}) & 0.05 / 0.05 / 2 / 3 / 6 & Case-dependent arrest
saturations; $\chi$ and $m$ kept within plausible ranges \\
\addlinespace
$S_c$, $q$ (immobilization function, Eq.~\ref{eq:model-f1}) & 0.65 / 1 &
Response midpoint of the switch-time experiment
\cite{jaiser.etal2016InvestigationFilm} \\
\addlinespace
$\alpha$ (calendering operator) & 0.5 & Surface reading \\
\addlinespace
Composition, $\varepsilon$, $H_0$, $\dot m$ & Literature values & Not
fitted \\
\bottomrule
\end{tabular}
\end{table}

$D_0$ is an effective diffusivity and not a material constant of PVDF in
NMP, because it absorbs non-equilibrium effects and the solution
properties, liquid viscosity and surface tension, that are set to water
values. The difference caused by the water values of the solvent constants
is likewise absorbed in the threshold, as a case-independent factor in
$\chi$. $\alpha$ is a coefficient of an observation operator and changes no
simulated field (\S\ref{sec:model-obs}).

The two parameters of the immobilization function, $(S_c, q)$, are not
determined individually. Within the plausible range they substitute for
each other almost completely, so what the switch-time experiment actually
determines is one number, the saturation at which binder motion ends
(\S\ref{sec:results-arrest}).

What was not fitted is the drying history of the film. The film thickness
during consolidation and the saturation during pore emptying are set by
the composition and the prescribed evaporative flux alone, and they were
never fitted to either schedule experiment. The time at which film
consolidation ends and the total drying time are therefore predictions in
both experiments, and the three-stage experiment was not used in any
calibration. The comparisons with dried films, and the midpoint and shape
of the switch-time response, are by contrast comparisons with the very
data the parameters were fitted to. Section~\ref{sec:results} states this
distinction for each comparison.

%% file: sections/04_results.tex
\section{Results}
\label{sec:results}

All comparisons in this section were computed with the single parameter
set of Table~\ref{tab:data-ledger}, and nothing was re-fitted between
subsections. Section~\ref{sec:results-endpoint} treats the binder
distribution of dried films, and Section~\ref{sec:results-schedule} the
interface binder against the time at which the drying rate is lowered.
Section~\ref{sec:results-arrest} treats the saturation at which the binder
becomes fixed, which the two comparisons determine together, and
Section~\ref{sec:results-ablation} shows which comparison fails when each
of the two arrest mechanisms is removed, and Section~\ref{sec:results-hlh}
treats the three-stage schedules, to which nothing was fitted.
Table~\ref{tab:results-comparison} collects the values compared between
model and measurement in this section, together with whether
each is a comparison with the data the parameters were fitted to
(calibration) or a comparison without fitting (prediction).

{\linespread{1}\selectfont\footnotesize
\setlength{\tabcolsep}{3pt}
\begin{longtable}{>{\raggedright\arraybackslash}p{1.9cm}>{\raggedright\arraybackslash}p{4.4cm}>{\raggedright\arraybackslash}p{1.9cm}>{\raggedright\arraybackslash}p{2.1cm}>{\raggedright\arraybackslash}p{2.3cm}}
\caption{Values compared between model and measurement}
\label{tab:results-comparison}\\
\toprule
Experiment & Quantity compared & Measured & Model & Status \\
\midrule
\endfirsthead
\multicolumn{5}{l}{Table~\ref{tab:results-comparison} (continued)}\\
\toprule
Experiment & Quantity compared & Measured & Model & Status \\
\midrule
\endhead
\bottomrule
\endlastfoot
Drying-rate series \cite{muller.etal2017InvestigationBinder} &
Rate sensitivity of the surface enrichment, $k$ (m$^2$\,s\,g$^{-1}$) &
0.203--0.212 & 0.211 & Calibration ($D_0$) \\
& Largest deviation of the four model points from a straight line &
Error bars $\approx\pm$0.08 & 0.004 & Prediction \\
\addlinespace
A400 series \cite{muller.etal2017InvestigationBinder} &
Model-to-measured ratio of $k$ between the two films & -- & 1.01 &
Calibration \\
& Thickness exponent $n$ & 0.314 & 0.317 & Calibration \\
& Surface enrichment (surface-to-interface ratio), slow 430 / fast
450~\textmu m & 1.155 (3.2) / 1.820 (6.2) & 1.150 (3.2) / 1.820 (6.2) &
Calibration \\
& RMS difference of normalized concentration, 17 interior blocks & -- &
0.091 & Calibration \\
& Enrichment-rate ratio (surface-side / deep-side half), fast
450~\textmu m & 1.12--1.49 & 1.44 & Calibration \\
& Same, slow 430~\textmu m & 0.74 & 1.20 & Calibration target, not
reproduced \\
& Interface concentration, slow 430~\textmu m & 0.84--0.95 & 0.858 &
Calibration \\
& Interface concentration, fast 450~\textmu m & 0.69--0.78 & 0.687 &
Calibration (0.5\% below the lower bound) \\
& Interface deficit (fraction of interface binder lost by the fast film) &
0.08--0.21 & 0.200 & Calibration \\
\addlinespace
Switch-time experiment \cite{jaiser.etal2016InvestigationFilm} &
Time at which film consolidation ends & 28 s ($t_{c,1}$) & 27.9 s &
Prediction \\
& Drying time of the film dried fast throughout & 65 s & 67 s &
Prediction \\
& Response midpoint $t_{50}$ & 35.9 s & 35.6 s & Calibration ($S_c$) \\
& RMS difference over the whole response & -- & 0.140 & Calibration \\
& Range over which the normalized interface level falls from 1 to 0
(10--90\%) & 28--46 s & 31--39 s & Shape of a calibration target, not
reproduced \\
& Change of interface binder of the film dried fast throughout (relative
scale) & $-14.9\%$ & $-6.3\%$ & Prediction, not reproduced \\
\addlinespace
\cite{jaiser.etal2016InvestigationFilm,jaiser.etal2017DevelopmentThreestage}
& Saturation below which the binder no longer moves & 0.55--0.64
(inferred from the experiments) & 0.65 ($S_c$, where $f_1$ reaches zero)
& Calibration result compared with the inference \\
\addlinespace
Three-stage experiment \cite{jaiser.etal2017DevelopmentThreestage} &
Drying time of the film dried slowly throughout & 162 s (calculated by the
source) & 157 s & Prediction \\
& Drying times of scenarios S1--S4 & 101.6, 94.8, 90.7, 86.8 s (calculated
by the source) & 100.7, 93.3, 88.9, 85.0 s & Prediction \\
& Interface level where quality is preserved (S1, S2) & 1.04, 0.92 &
1.00, 1.00 & Prediction \\
& Interface level after a later switch (S3, S4) & 0.67, 0.74 & 0.97, 0.85 &
Prediction; the drop comes at a later switch than measured \\
\end{longtable}
}

\subsection{Final binder distribution across drying rate and film thickness}
\label{sec:results-endpoint}

A single parameter set, that of Table~\ref{tab:data-ledger}, reproduces the
binder distribution after drying in films that differ in both drying rate
and thickness. The comparison is with the two series of
\cite{muller.etal2017InvestigationBinder}: the drying-rate series, in which
films about 70~\textmu m thick were dried at four drying rates between 0.5
and 2.8~g\,m$^{-2}$\,s$^{-1}$, and two cross-sections of films six times
thicker, 430 and 450~\textmu m, dried at 1.0 and 2.8~g\,m$^{-2}$\,s$^{-1}$.
This agreement is not an independent prediction but the result of
calibration: the effective binder diffusivity prefactor $D_0$ was fitted to
the rate sensitivity of the drying-rate series, and the coefficients of the
corrective terms mainly to the two cross-sections
(Table~\ref{tab:data-ledger}). What this comparison provides is therefore
less the agreement itself than the shape of binder transport that appears
when the two series are fitted with one set of coefficients. The reference
case is the base model, with all corrective terms removed and only $D_0$
re-fitted to the same rate sensitivity
($D_0 = 3.95\times10^{-10}$~m$^2$\,s$^{-1}$), which fits the thin films but
departs from the thick ones.

\begin{figure}
  \centering
  \includegraphics[width=\linewidth]{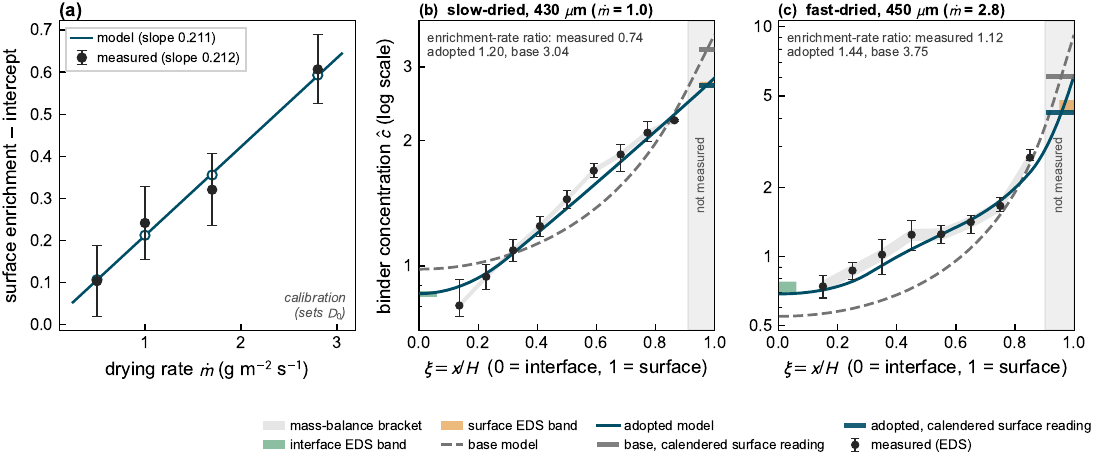}
  \caption{Comparison of the model with the endpoint measurements of
\cite{muller.etal2017InvestigationBinder}: (a) surface enrichment of the
70~\textmu m drying-rate series (rate sensitivity only); (b, c) binder
concentration profiles of the slow-dried (430~\textmu m) and fast-dried
(450~\textmu m) films on the mass-balance absolute scale (logarithmic
axis), with the base model for comparison. Experimental data digitized
from the source, Figs.~2 and 4.}
  \label{fig:results-endpoint}
\end{figure}

In the thin films, faster drying moves more binder toward the surface, and
the surface-to-interface concentration ratio grows exponentially with
drying rate. Fig.~\ref{fig:results-endpoint}a plots the logarithm of this
ratio, the surface enrichment, on the vertical axis, and the four measured
points lie on one straight line against drying rate. The rate sensitivity
of the model, $k = 0.211$, lies within the measured 0.203--0.212: the
surface-to-interface ratio grows by about 23\% for every
1~g\,m$^{-2}$\,s$^{-1}$ increase in drying rate. Because $D_0$ was set by
this value, the only feature not fitted here is the linearity, and the four
model points deviate from a straight line by at most 0.004, far less than
the measured error bars (about $\pm$0.08 in surface enrichment).
Fig.~\ref{fig:results-endpoint}a compares only the slope and leaves out the
height of the line, its intercept. The source measured the surface value by
EDS on the top face of the film and the interface value on the bottom face
exposed by peeling off the current-collector foil, so the binder appears
differently on the two faces. Measurements of samples of nearly the same
composition by both XPS and EDS in the source supplement show that the
height of the line changes strongly with the measurement method while the
slope stays within 0.20--0.25 (\S\ref{sec:model-obs}). The measured range
0.203--0.212 consists of the XPS value and the value of the EDS series of
Fig.~\ref{fig:results-endpoint}a.

As the film becomes thicker, the effect of drying rate on binder migration
also grows, but not in proportion to thickness. The rate sensitivity of the two 400~\textmu m
films is obtained by dividing the difference of their surface enrichments
by the difference of their drying rates, 1.8~g\,m$^{-2}$\,s$^{-1}$. The
measured value is about 1.75 times that of the 70~\textmu m series, whereas
the film is about six times thicker (mean dry thickness 440~\textmu m
against 73.5~\textmu m). The rate sensitivity thus grows as the 0.314 power
of thickness (thickness exponent $n = 0.314$). In the quasi-steady limit, in
which the surface enrichment follows the P\'eclet number, proportional to
the product of film thickness and drying rate, the rate sensitivity would
grow with thickness by the same factor of six. The adopted model gives a
factor of 1.76 ($n = 0.317$; model-to-measured ratio of the rate
sensitivity between the two films, 1.01), and the surface enrichments of
the two films, 1.150 and 1.820 (surface-to-interface ratios 3.2 and 6.2),
equal the measured 1.155 and 1.820. The base model gives a factor of 3.2
($n = 0.64$), exaggerating the effect of drying rate in the thick films to
about 1.8 times the measured value. This weaker-than-proportional growth
arises from the way the binder stops during pore emptying. Removing both
arrest mechanisms while keeping the other corrective terms returns $n$ to
0.644, the value of the base model; keeping only the immobilization
function gives 0.383, and keeping only the threshold 0.289
(\S\ref{sec:results-ablation}). The threshold has this large effect because
its position shifts with the P\'eclet number. In a thick film dried fast,
the surface layer becomes strongly concentrated during consolidation and
the liquid viscosity rises, so liquid pumping fails as soon as pore emptying
begins and the binder stops (\S\ref{sec:model-corr-threshold},
\S\ref{sec:model-corr-roles}). The larger the thickness and the drying rate
together, the shorter the time left for the enrichment to grow during pore
emptying.

Inside the thick films, the binder is moved toward the surface at nearly
the same speed at every depth. Figs.~\ref{fig:results-endpoint}b,c plot the
binder concentration from the interface (left, $\xi = 0$) to the surface
(right, $\xi = 1$) on a logarithmic scale, so a constant depth enrichment
rate appears as a straight line. The gray region at the right end is the
top block that the source excluded from its analysis, and the dark bar above it is the
surface reading of the model, the average over the top layer mixed by
calendering (\S\ref{sec:model-obs}), compared with the orange band, the
measured surface EDS range. The green band at the left end is the measured
interface EDS range. The measured block averages form nearly straight lines
in both films. Because the depth enrichment rate is proportional to the
advection velocity that moves the binder upward
(\S\ref{sec:model-corr-pumping}), a straight line means that this velocity
is nearly the same at every depth. The adopted model follows this shape in
the fast-dried film: the enrichment-rate ratio, the mean enrichment rate of
the surface-side half divided by that of the deep-side half, is 1.44,
within the measured 1.12--1.49. The base model bends upward near the
surface, and its ratio reaches 3.8, because its advection velocity grows in
proportion to height and the binder moves mainly near the surface. In the
slow-dried film the adopted model is not as straight as the measurement
either (enrichment-rate ratio 1.20, measured 0.74, base model 3.0); the
enrichment-rate ratio of this film was not reproduced although its
cross-section was used for calibration.

The amount of interface binder, which governs adhesion, also agrees on the
absolute scale. The measured cross-sections are relative values per block,
so the absolute scale was set such that the total binder in the film is
conserved. The scale factor of each cross-section was chosen to minimize
the difference from the interior blocks (nine for the slow-dried film,
eight for the fast-dried one) within the range that conservation of the
total allows for the two end blocks excluded from the analysis (light gray band in
Figs.~\ref{fig:results-endpoint}b,c). On this scale the fast-dried film
loses 20\% of the interface binder relative to the slow-dried film, against
the measured 8--21\%. The interface concentration is 0.858 in the
slow-dried film (measured 0.84--0.95) and 0.687 in the fast-dried film; the
latter is 0.5\% below the lower bound of the measured range, 0.69, and is
reported as such. Over all 17 interior blocks of the two cross-sections,
the RMS difference between model and measurement, in concentration
normalized by the initial concentration, is 0.091. The base model leaves
the interface of the slow-dried film almost undepleted at 0.98 and depletes
that of the fast-dried film to 0.55, so the exaggerated rate sensitivity of
the thick films appears directly at the interface.

In summary, one parameter set reproduces, for films that differ by a
factor of six in thickness and a factor of 5.6 in drying rate (P\'eclet
numbers 0.19--6.5), how strongly the binder accumulates at the surface and
how this depends on thickness, the shape of the distribution with depth,
and the amount left at the interface. Fitting these shapes together
requires different transport mechanisms: the localized pumping that makes
the consolidation convection plug-like, for the depth-uniform enrichment
rate (\S\ref{sec:model-corr-pumping}); the carrier coupling, for the
interface level of the fast-dried film (\S\ref{sec:model-corr-coupling});
the dual-mode shrinkage, for the surface spike and the absolute interface
level (\S\ref{sec:model-corr-shrinkage}); and the two arrest mechanisms,
for the weaker-than-proportional growth with thickness. Because the
enrichment-rate ratio of the slow-dried film was not reproduced (model
1.20, measured 0.74), the picture of depth-uniform advection is confirmed
for the fast-dried film. A dried film, moreover, records only where the
binder finally came to rest, not when during drying it stopped moving.
That time is determined by the switch-time experiment of the next section.

\subsection{Interface binder against the time of the drying-rate switch}
\label{sec:results-schedule}

When the drying rate is lowered once during drying, the interface binder
depends on when the rate is lowered, and the model follows this
dependence. The comparison is with the switch-time experiment of
\cite{jaiser.etal2016InvestigationFilm}, in which films about
78~\textmu m thick were dried starting at 1.19~g\,m$^{-2}$\,s$^{-1}$, the
rate was lowered to 0.52~g\,m$^{-2}$\,s$^{-1}$ at the switch time
$t_{\mathrm{trans}}$, and the final interface binder and adhesion were
measured for a range of switch times. The model computes the interface
binder concentration, not peel strength, and the two are connected only
through the monotonic relation measured in
\cite{jaiser.etal2016InvestigationFilm}: over this range, peel strength
increases monotonically with the binder concentration at the fracture
plane. Comparisons with adhesion are therefore valid for ordering and
timing only. In the measurement, the switch time
changes the outcome only within a short window of the drying: the adhesion
equals that of a film dried slowly throughout if the rate is lowered before
28~s, and that of a film dried fast throughout if it is lowered after 44~s.
The model explains this window as the interval from the moment film
consolidation ends to the moment the binder stops moving. The start of the
window is predicted without fitting anything to this experiment, and its
central position is the result of fitting the binder immobilization
function to this experiment.

\begin{figure}
  \centering
  \includegraphics[width=\linewidth]{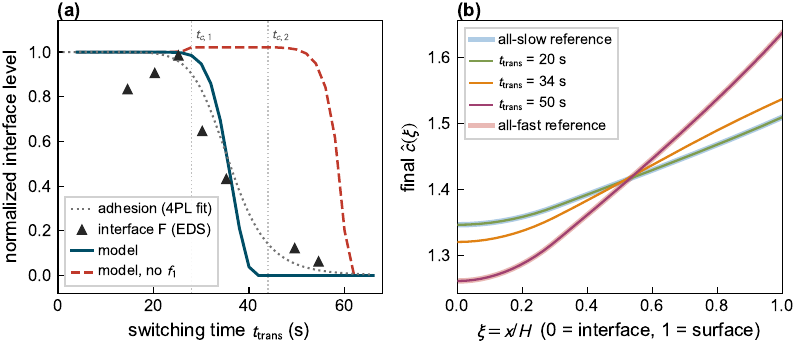}
  \caption{Comparison of the model with the switch-time experiment of
\cite{jaiser.etal2016InvestigationFilm} (78~\textmu m film, 1.19 to
0.52~g\,m$^{-2}$\,s$^{-1}$ at $t_{\mathrm{trans}}$): (a) normalized
interface level against switch time (dark solid line: model; red dashed
line: model without the immobilization function; triangles: interface
fluorine EDS, fluorine marking the PVDF binder; gray dotted curve:
four-parameter logistic fit of the adhesion reported in the source; gray
vertical dotted lines: the two characteristic times reported in the source,
$t_{c,1} = 28$~s and $t_{c,2} = 44$~s); (b) final binder profiles at three
switch times and the two reference dryings. Experimental data digitized
from the source, Fig.~3.}
  \label{fig:results-schedule}
\end{figure}

Fig.~\ref{fig:results-schedule}a plots the normalized interface level, on a
scale with the interface binder of the film dried slowly throughout at 1
and that of the film dried fast throughout at 0. Model and measurement are
each placed relative to their own two reference dryings, so the absolute
amount of interface binder lost by the film dried fast throughout does not
appear on this scale (Table~\ref{tab:results-comparison}).
Fig.~\ref{fig:results-schedule}b shows the final binder distribution at
each switch time together with those of the two reference films.

The start of the window is the time at which film consolidation ends, and
the model predicts it with no parameter fitted to this experiment. The
drying history of the model, the film thickness during consolidation and
the saturation during pore emptying, is set by the film composition and the
prescribed evaporative flux alone (\S\ref{sec:data-ledger}). In that
history the film stops shrinking and its pores begin to empty at 27.9~s,
equal to the measured $t_{c,1} = 28$~s. With the same history, the film
dried fast throughout reaches the final moisture content at 67~s, against
the 65~s reported by the source. Lowering the rate before consolidation
ends leaves no trace of the high-rate stage: the final distribution of the
model then equals that of a film dried slowly from the start to within
$10^{-5}$ in normalized concentration (the $t_{\mathrm{trans}} = 20$~s curve
in Fig.~\ref{fig:results-schedule}b). A film switched after the window
likewise equals the film dried fast throughout (the
$t_{\mathrm{trans}} = 50$~s curve).

Between these times, the later the switch, the less interface binder
remains. At the end of consolidation the film enters pore emptying with a
distribution in which the binder has already accumulated toward the
surface. If the subsequent drying is slow, diffusion returns that
enrichment toward the interface before the binder stops; if it is fast,
the enrichment is fixed nearly in place. The later the switch, the shorter
the time for this return, so the $t_{\mathrm{trans}} = 34$~s curve,
switched inside the window, lies between the two reference curves
(Fig.~\ref{fig:results-schedule}b). This response starts from the
distribution left by consolidation and is formed during pore emptying. It
therefore cannot be reproduced by a model that ends the calculation with
consolidation or by one that treats pore emptying alone; the two stages
must be connected in one calculation.

The central position of the window is set by the binder immobilization
function, and this part is calibration. With the arrest point $S_c$ fitted
to the midpoint of this response, the model midpoint is 35.6~s, equal to
the measured 35.9~s, and the RMS difference over the whole response is
0.140 (the model values were obtained by interpolating calculations at
twelve switch times between 4 and 64~s). Without the immobilization
function the binder keeps moving until liquid pumping fails at about 60~s,
and the midpoint moves to 59.1~s, outside the measured window (red dashed
line in Fig.~\ref{fig:results-schedule}a). This experiment thus requires
the binder to stop early in pore emptying, well before liquid pumping
fails. The saturation at which it stops is treated in
\S\ref{sec:results-arrest}.

The response of the model is, however, steeper than the measured one. The
measured interface level falls from 1 to 0 over about 28--46~s, a width of
18~s, whereas the model does so over 31--39~s, about 8~s: the centre agrees
but the width is about half. This narrow window is a limitation of the
model, and it appears again in the three-stage experiment
(\S\ref{sec:results-hlh}). Separately, the model underestimates how much
interface binder the film dried fast throughout loses
(Table~\ref{tab:results-comparison}). Both are discussed in
\S\ref{sec:discussion-limits}.

\subsection{The arrest point}
\label{sec:results-arrest}

What the switch-time experiment actually determines is one number, the
saturation at which binder motion ends during pore emptying, and the value
obtained by the model is consistent with the inferred fixation range derived
from the source experiments. There is only one number because the two
parameters of the immobilization function substitute for each other.
Changing the pair of arrest point and immobilization exponent, $(S_c, q)$,
from the adopted $(0.65, 1)$ to $(0.60, 2)$, $(0.70, 1)$ or $(0.55, 3)$
fits the dried-film comparison equally well and moves the response midpoint
$t_{50}$ by only a few seconds; across these pairs the saturation at which
the binder mobility reaches zero ranges from 0.55 to 0.70.

Read from the values reported by the source experiments, the binder no
longer moves below saturations of 0.55--0.64. Both sources converted the
characteristic times of their switch-time experiments to solvent volume
fractions. The second of these, after which a change of drying rate no
longer changed the interface, is 0.442 in
\cite{jaiser.etal2016InvestigationFilm}, where it corresponds to
$t_{c,2} = 44$~s, and 0.48 in \cite{jaiser.etal2017DevelopmentThreestage},
which names it the binder immobilization point. Converted to saturation
with the definition of the liquid volume fraction used in these sources
(\S\ref{sec:data-sets}), they give 0.55 and 0.64, the ends of the inferred
fixation range. The range is not a direct measurement of fixation but an
inference from the observation that changing the drying rate after that
point no longer changed the interface.

The calibrated immobilization function reaches zero at $S_c = 0.65$, just
above this range. The 0.64 of \cite{jaiser.etal2017DevelopmentThreestage},
obtained from a switch-time experiment that was not used in any
calibration, differs from $S_c$ by 0.01. Against the lower end, the 0.55 of
\cite{jaiser.etal2016InvestigationFilm}, the model stops the binder earlier
by about 0.1 in saturation. This is the narrow window of
\S\ref{sec:results-schedule} seen on the saturation scale: the response of
the model ends at 39~s, whereas the measured response ends at 46~s. The
reading that uses no binder parameter is the midpoint. Converting the
measured midpoint $t_{50} = 35.9$~s through the drying history of the
model, the time course of saturation that was not fitted to this
experiment (\S\ref{sec:results-schedule}), gives a saturation of 0.79, and
the calibrated function halves the binder mobility at a similar saturation,
$S = 0.825$. In this picture the binder slows down from the moment the pores
begin to empty, moves at half its mobility near $S = 0.8$, and has stopped
by $S \approx$ 0.55--0.65. No number was adjusted toward another. This
agreement confirms that the picture of the binder stopping in the first part
of pore emptying is consistent, but it does not prove the mechanism.

\subsection{Removal of each arrest mechanism}
\label{sec:results-ablation}

The two arrest mechanisms explain different experiments, and measurements
on dried films alone therefore cannot determine when the binder stops. To
show this, the two mechanisms were removed separately and together. In
each configuration only $D_0$ was re-fitted to the rate sensitivity of the
drying-rate series, and the other parameters were kept at the adopted
values. Table~\ref{tab:results-ablation} lists the results.

\begin{table}
\caption{Comparison with each arrest mechanism removed (immobilization
function: binder immobilization function $f_1$, Eq.~\eqref{eq:model-f1};
threshold: dual-porosity pumping-failure threshold,
Eq.~\eqref{eq:model-sfail}; in each configuration only $D_0$ re-fitted to
the drying-rate series; measured values in parentheses; bold marks values
outside the measured range)}
\label{tab:results-ablation}
\centering
\footnotesize
\setlength{\tabcolsep}{3pt}
\resizebox{\textwidth}{!}{%
\begin{tabular}{lccccccccc}
\toprule
Configuration & $D_0$ & \begin{tabular}[c]{@{}c@{}}Deviation\\from line\\($\pm$0.08)\end{tabular} & \begin{tabular}[c]{@{}c@{}}Ratio of\\$k$ (1)\end{tabular} & \begin{tabular}[c]{@{}c@{}}$n$\\(0.314)\end{tabular} & \begin{tabular}[c]{@{}c@{}}Surface enr.,\\fast (1.820)\end{tabular} & \begin{tabular}[c]{@{}c@{}}$\hat c_i$, fast\\(0.69--0.78)\end{tabular} & \begin{tabular}[c]{@{}c@{}}Interface\\deficit\\(0.08--0.21)\end{tabular} & \begin{tabular}[c]{@{}c@{}}$t_{50}$\\(35.9 s)\end{tabular} & \begin{tabular}[c]{@{}c@{}}Response\\RMS\end{tabular} \\
\midrule
Both present (adopted) & $2.52\times10^{-10}$ & 0.004 & 1.01 & 0.317 & 1.820 & \textbf{0.687} & 0.200 & 35.6 & 0.140 \\
No immobilization function & $2.70\times10^{-10}$ & 0.022 & 0.96 & 0.289 & 1.784 & 0.694 & 0.193 & \textbf{59.1} & \textbf{0.521} \\
No threshold & $2.52\times10^{-10}$ & 0.004 & \textbf{1.13} & \textbf{0.383} & \textbf{1.904} & \textbf{0.604} & \textbf{0.296} & 35.6 & 0.140 \\
Neither & $4.06\times10^{-10}$ & 0.015 & \textbf{1.81} & \textbf{0.644} & \textbf{2.375} & \textbf{0.437} & \textbf{0.500} & \textbf{63.6} & \textbf{0.548} \\
\bottomrule
\end{tabular}}
\end{table}

Removing the immobilization function leaves the dried-film comparison
nearly unchanged and breaks only the switch-time response. The endpoint
values stay close to those of the adopted model, close enough that a study
looking only at endpoints would have accepted this variant. The response
midpoint, however, moves from 35.6~s to 59.1~s, outside the measured
window, because the binder keeps moving until liquid pumping fails at about
60~s (\S\ref{sec:results-schedule}).

Removing the pumping-failure threshold does the opposite. The switch-time
response does not change to the reported decimal, while the endpoints of
the fast-dried thick film move outside the measured ranges: the
model-to-measured ratio of the rate sensitivity between the two films
becomes 1.13, the thickness exponent 0.383, the interface concentration of
the fast-dried film 0.604 and the interface deficit 0.296. Removing both
degrades every comparison.

Each mechanism leaves the other's experiment unaffected because the two act
at different saturations. In the fast-dried 450~\textmu m film the carrier
network fails at $S_{\mathrm{fail}} \approx 1$, where pore emptying begins,
and because $f_1 \approx 1$ at that saturation, only the threshold can
produce this arrest. In the 78~\textmu m film of the switch-time experiment
the threshold is at $S_{\mathrm{fail}} = 0.15$--$0.18$, but the
immobilization function has already reduced the mobility to zero much
earlier, below $S \approx 0.7$, so the threshold has no effect and the
response is set by the immobilization function alone. This is the
quantitative content of the separation of roles stated in
\S\ref{sec:model-corr-roles}.

The immobilization function is also consistent with the dried-film
comparison. The coefficients of the corrective terms were fitted to the
dried films before the immobilization function was introduced, and the
immobilization function was set by the switch-time experiment alone. Yet
introducing it moved the endpoint comparison toward the measurements: the
deviation of the drying-rate series from a straight line changed from
0.022 to 0.004, the surface enrichment of the fast-dried 450~\textmu m film
from 1.784 to the measured 1.820, and the thickness exponent from 0.289 to
0.317.

In summary, the immobilization function could not be obtained from the
drying-rate series and the two cross-sections, however well they were
reproduced; it entered the model only when the switch-time experiment
required it, and its one effective parameter was set by that experiment.
Determining when the binder stops requires an experiment that changes the
drying rate during drying.

\subsection{The three-stage experiment}
\label{sec:results-hlh}

In the unfitted prediction of the three-stage schedules, the model
reproduces the drying times of all four scenarios and the full recovery of
the interface when the drying rate is lowered early, whereas the drop of
the interface after a later switch comes at a later switching point than
measured. The comparison is with the three-stage experiment of
\cite{jaiser.etal2017DevelopmentThreestage}, in which films with a dry
loading of 72~g\,m$^{-2}$ were dried at the high rate, switched to the low
rate when the liquid volume fraction reached $\phi_{t1}$, and returned to
the high rate at $\phi_{t2} = 0.45$. The design passes the beginning of
drying, before binder transport has built a damaging gradient, and the end,
after the binder is fixed, at the high rate, and the window in which the
interface is decided at the low rate. Adhesion was measured in four
scenarios with $\phi_{t1}$ = 0.63, 0.60, 0.58 and 0.56 and is compared with
the interface level through the monotonic relation of
\S\ref{sec:results-schedule}; the drying time of each scenario was
calculated by the source from the measured drying rates. Nothing was fitted
to this experiment: every parameter is from the set of
Table~\ref{tab:data-ledger}, and the schedule enters only through the
prescribed evaporative flux $\dot m(t)$.

\begin{figure}
  \centering
  \includegraphics[width=\linewidth]{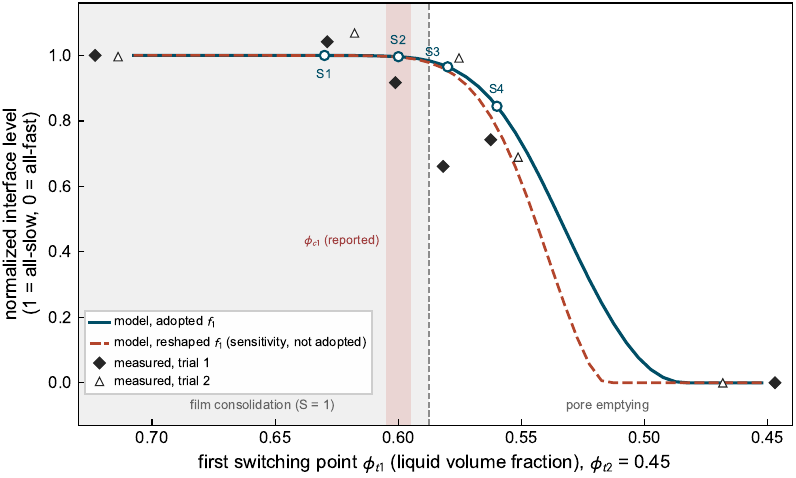}
  \caption{Comparison of the predicted interface level with the normalized
adhesion of the three-stage schedules of
\cite{jaiser.etal2017DevelopmentThreestage} over the first switching point
$\phi_{t1}$ at $\phi_{t2}$ = 0.45, with a non-adopted variant of the
immobilization function. Experimental data digitized from the source,
Fig.~5.}
  \label{fig:results-hlh}
\end{figure}

The horizontal axis of Fig.~\ref{fig:results-hlh} gives the first switching
point as the liquid volume fraction $\phi$, the volume of liquid divided by
the volume of liquid and solids (\S\ref{sec:data-sets}). Because $\phi$
decreases as drying proceeds, later points lie to the right. During film
consolidation the pores are filled with solvent ($S = 1$) and $\phi$ falls
as the film shrinks; from the end of consolidation in the model at
$\phi \approx 0.59$ (gray dashed line, with gray shading to its left) the
pores empty and $\phi$ falls with the saturation. The vertical axis is the
interface level, 1 for drying at the low rate throughout and 0 for the high
rate throughout. The pink band marks the liquid volume fraction that
\cite{jaiser.etal2017DevelopmentThreestage} reports for the end of film
shrinkage, $\phi_{c1} \approx 0.60$, which corresponds to $t_{c,1} = 28$~s
of the switch-time experiment. The dark blue solid curve is the adopted
model and the red dashed curve a variant with a different shape of the
immobilization function ($S_c = 0.74$, $q = 0.6$), not adopted and shown
for sensitivity; the open circles mark the four scenarios, and the black
diamonds and open triangles are the normalized adhesion of the two coating
trials of the source.

The drying times follow the source without fitting. The film dried slowly
throughout dries in 157~s in the model and in 162~s according to the
source, within 4\%, and the four scenarios dry in 100.7, 93.3, 88.9 and
85.0~s in the model against 101.6, 94.8, 90.7 and 86.8~s calculated by the
source, within 2~s each. For scenario S2, the drying is shortened relative
to the film dried slowly throughout by about 40\% in both. These times test
the drying history of the model, which is set by the film composition and
the prescribed evaporative flux alone.

The return to the high rate has no effect on the interface in these
schedules, because it comes after the binder has stopped. The return point
$\phi_{t2} = 0.45$ corresponds to a saturation of 0.57
(\S\ref{sec:data-sets}), below $S_c = 0.65$, where the immobilization
function has already reduced the binder mobility to zero
(Fig.~\ref{fig:model-arrestmap}a), and the source placed the return there
deliberately, at about its second characteristic solvent volume fraction.
In the model, each three-stage scenario therefore gives the same interface
as a single switch from the high to the low rate at the same time, to
within $10^{-4}$ in normalized interface level. The three-stage experiment
thus measures the same window as the switch-time experiment of
\S\ref{sec:results-schedule}, on a second set of films and with the switch
placed by liquid volume fraction instead of time.

Where the drying rate is lowered early, the interface recovers fully, as
measured. In scenarios S1 and S2, switched before the end of film
consolidation, the model gives 1.00 for both, against the measured 1.04 and
0.92; as in the switch-time experiment, a switch made before the pores
begin to empty leaves no trace of the high-rate stage. After a later switch
the interface drops in both model and measurement, and at the late end of
the family, where the switch is delayed to $\phi_{t2} = 0.45$ and the
low-rate stage vanishes, both reach the level of the film dried fast
throughout. The drop of the model, however, comes at a later switch than
measured: S3 and S4 give 0.97 and 0.85, against the measured 0.67 and 0.74.
This is the narrow window of \S\ref{sec:results-schedule}. On the time
scale of the model, S3 and S4 switch at 29.8 and 32.8~s, where the model
response has only begun to fall, whereas the broader measured response of
the switch-time experiment has already fallen to 0.84 and 0.69 at the same
times. The measured values of S3 and S4 are themselves not in order, and the
second coating trial gives 0.99 at nearly the same switching point as S3
($\phi_{t1}$ = 0.575 against 0.58), so the scatter between trials is as
large as the difference between model and measurement at these two points.
This limitation is discussed in \S\ref{sec:discussion-limits}.

%% file: sections/05_discussion.tex
\section{Discussion}
\label{sec:discussion}

\subsection{Placement of the low-rate stage}
\label{sec:discussion-placement}

What decides the interface is not the average drying rate but where along
the saturation history the drying intensity is placed. The interface
responds to the drying rate only in the window from the end of film
consolidation to the moment the binder stops
(\S\ref{sec:results-schedule}, \S\ref{sec:results-arrest}), so a schedule
that slows down only inside that window obtains most of the adhesion
benefit at the cost of a small part of the drying time. The model is
intended as the electrode-side model for schedule optimization in
subsequent work; this paper is limited to showing that its predictions are
quantitatively reliable.

\begin{figure}
  \centering
  \includegraphics[width=\linewidth]{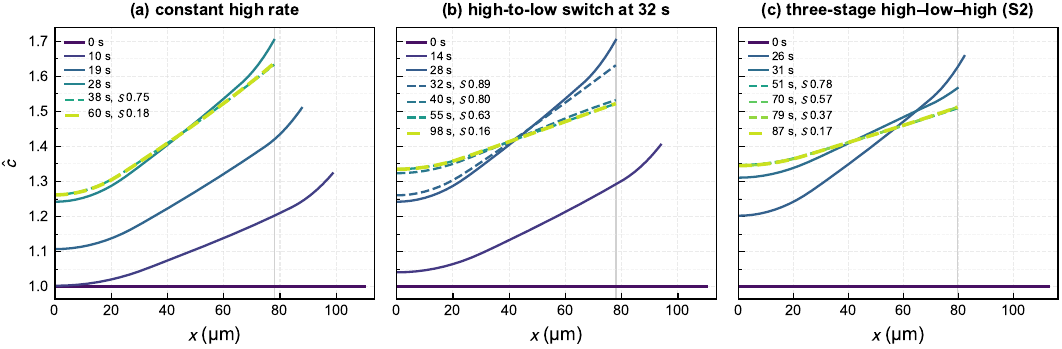}
  \caption{Evolution of the dissolved binder profile on the dimensional
coordinate at frozen parameters: (a) constant high rate; (b) high-to-low
switch at 32~s; (c) three-stage high--low--high schedule S2.}
  \label{fig:process-evolution}
\end{figure}

Fig.~\ref{fig:process-evolution} shows how this sensitivity to placement
arises, as the change of the binder distribution during drying. The panels
plot the binder concentration $\hat c(x)$ at successive times on the
dimensional coordinate for (a) the constant high rate, (b) the high-to-low
switch of \S\ref{sec:results-schedule} and (c) the three-stage schedule S2.
Solid curves belong to film consolidation and dashed curves to pore
emptying, with the saturation given in the legend. During consolidation the
domain shortens while the whole profile rises, because the binder
inventory $\int \hat c\,dx$ is conserved as solvent leaves. At the stage
transition the last consolidation profile is passed on unchanged. In pore
emptying the constant high rate stops the inherited enrichment nearly where
it stands, whereas the low-rate stages of (b) and (c) give diffusion time
to return the surface enrichment toward the interface before transport
stops.

\begin{figure}
  \centering
  \includegraphics[width=\linewidth]{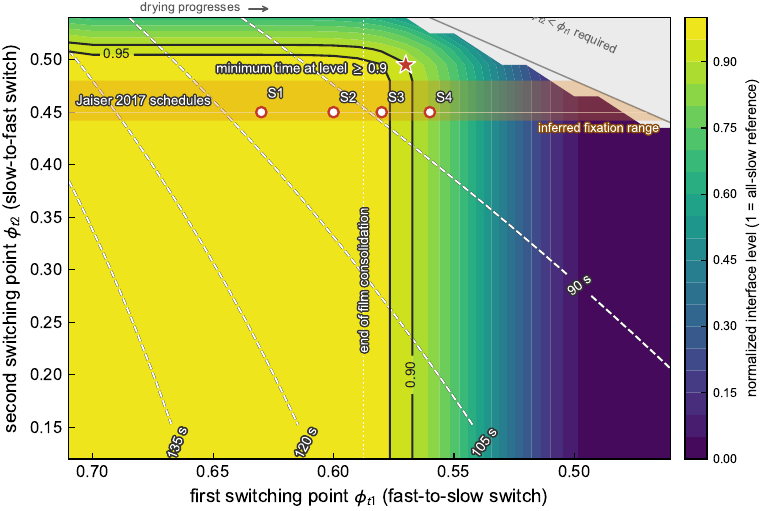}
  \caption{Schedule design map: predicted interface level over the
switching plane $(\phi_{t1}, \phi_{t2})$ at frozen parameters
(710 schedules), with the total drying time overlaid.}
  \label{fig:process-designmap}
\end{figure}

Fig.~\ref{fig:process-designmap} shows that extending the four published
scenarios to every placement of the low-rate stage reveals a region that
recovers nearly all of the interface while greatly shortening the drying
time. The two coordinates of the map are the liquid fraction $\phi_{t1}$
at which the low-rate stage begins and the liquid fraction $\phi_{t2}$ at
which it ends; since the liquid fraction decreases monotonically during
drying, smaller values of either coordinate are later. The colour gives the
interface level normalized between the films dried slowly (1) and fast (0)
throughout, white dashed contours give the total drying time, and black
contours mark interface levels 0.90 and 0.95. The star marks the shortest
schedule that keeps the interface level at or above 0.9, circles place the
four published scenarios (all at $\phi_{t2} = 0.45$), the orange band marks
the inferred fixation range $\phi_{t2}$ = 0.44--0.48, and the gray region
collects placements that are not in order
($\phi_{t2} \geq \phi_{t1} - 0.02$). The published scenarios lie along one
horizontal line of this plane, $\phi_{t2} = 0.45$, and in the model S1--S3
are inside the region where the interface level is 0.9 or higher. Inside
this region the drying time is still far shorter than that of the film
dried slowly throughout.

The map also shows that the first switching point decides the interface.
The region where the interface level is 0.9 or higher is bounded almost
entirely by the first switching point, at $\phi_{t1} \approx 0.57$, shortly
after the end of film consolidation, and hardly depends on the second
switching point below $\phi_{t2} \approx 0.50$. The reason is the one found
in \S\ref{sec:results-hlh}: once the saturation has fallen below $S_c$, the
binder no longer moves, so a return to the high rate after that point does
not change the interface. The return can therefore be brought forward from
the published $\phi_{t2} = 0.45$ to about 0.50 ($S \approx 0.70$) while the
interface level stays above 0.98 at $\phi_{t1} = 0.60$, and the shortest
schedule that keeps the interface level at or above 0.9 (star) dries in
81~s, against 157~s for the film dried slowly throughout. The edge at
$\phi_{t1} \approx 0.57$, however, inherits the narrow window of the model:
because the drop of the model comes at a later switch than measured
(\S\ref{sec:results-hlh}), the map places this edge later than the
measurements would, and the first switch should be made earlier than the
edge, around the end of film consolidation.

\subsection{Limitations and future work}
\label{sec:discussion-limits}

This study has two limitations. First, the interface of the model responds
to the drying schedule in the right direction but differs from the
measurements in two ways. In the switch-time experiment, the window over
which the interface responds to the switch time is narrower than measured:
its centre agrees, but it is about half as wide
(\S\ref{sec:results-schedule}), and the same narrow window places the drop
of the three-stage experiment at a later switch than measured
(\S\ref{sec:results-hlh}). Separately, the size of the interface change
that a schedule produces is smaller than measured for the film dried fast
throughout (\S\ref{sec:results-schedule},
Table~\ref{tab:results-comparison}). This interfacial response compression
underestimates the damage that fast drying does to the interface, and the
narrow window lets the model accept a later first switch than the
measurements allow, so in both respects the model is non-conservative for
process design. Changing the shape of the immobilization function did not
remove either difference, and their three candidate causes, an
arrest that is local rather than film-averaged, an arrest that is gradual
in time, and nonlinearity of the measured relative scale, cannot be
separated with the present data. Second, the comparisons with dried films
are mostly comparisons with the very data to which the corrective-term
coefficients were fitted. The unfitted comparisons are the values marked as
predictions in Table~\ref{tab:results-comparison}, and among them the
drying times of the two schedule experiments and the outcome of the
three-stage experiment test the drying history and the schedule response.

Identifying the causes requires more switch-time experiments of the type of
\cite{jaiser.etal2016InvestigationFilm}, with depth resolution. The
prescribed evaporative flux $\dot m(t)$ is exactly what the zone settings
of the oven produce. With experiments published together with their
air-side conditions, $\dot m(t)$ could be replaced by the drying-rate
history that the process model of Joo et
al.~\cite{joo.etal2026StatedependentTransient} computes from the zone
settings, so that oven energy and interface quality are evaluated in the
same calculation. The same two-stage chain and drying-rate interface can
also be carried over to the water-based anodes treated by that
electrode--process model. Dissolved CMC is a candidate for the same
framework, whereas SBR latex particles require a separate treatment of
particle transport, and the time at which the binder stops would again
have to be determined by switch-time experiments. Pore-scale models
\cite{weichel.etal2026PorescaleModeling} are candidates to replace the
corrective terms with resolved physics.

%% file: sections/06_conclusions.tex
\section{Conclusions}
\label{sec:conclusions}

In this study, a continuous simulation was developed that follows the
migration of dissolved binder during electrode drying across film
consolidation and pore emptying in one calculation. The simulation is
driven by a prescribed evaporative flux and closed by five corrective
terms, each introduced to reproduce a specific observation. From the
comparison of one parameter set, not re-fitted for each data set, with
three published data sets, it was found that the model reproduces the
response of the binder distribution to the three operating variables of
the drying oven together. For the drying-rate series it reproduces the
dependence of the surface-to-interface ratio on drying rate with a slope of
0.211 (measured 0.203--0.212); for films six times thicker it reproduces how the
effect of drying rate on binder migration grows less than in proportion to
thickness (thickness exponent 0.317, measured 0.314) and the depth distribution on an
absolute scale, and the P\'eclet numbers of the two series together range
over a factor of 34. For the switch-time experiment, the start of the
window in which the interface responds to the drying-rate switch was
predicted without fitting to coincide with the end of film consolidation
(27.9~s, measured 28~s), and the response midpoint fitted through the arrest
point is 35.6~s (measured 35.9~s). For the unfitted three-stage schedules,
the model reproduces the drying times of the four scenarios within 2~s of
those calculated by the source, a shortening of about 40\% in both, and the
full recovery of the interface when the drying rate is lowered early
(interface level 1.00, measured 0.92--1.04); the return to the high rate in
these schedules comes after the binder has stopped and does not change the
interface. Removing the two arrest mechanisms one at a time showed that
measurements on dried films alone cannot determine when the binder stops.
That time is determined only by the switch-time experiment, and the arrest
point of the calibrated immobilization function, $S$ = 0.65, is just above
the fixation range inferred independently of the model, 0.55--0.64.

Therefore, the model can be used as the electrode-side model in the design
of drying schedules for NMP-based electrodes, and, combined with a
coater-scale process model through the shared drying-rate interface, it is
expected to evaluate oven energy and interface quality in the same
calculation. The main remaining limitations are that the interface responds
to the switch time over a window about half as wide as measured and that
the film dried fast throughout loses less interface binder than measured;
identifying their causes requires measurements of the binder distribution
by depth during drying. Extending the same approach to water-based anodes,
to connect it with the earlier electrode--process model for such anodes,
is the next step.